\documentclass[lettersize,journal]{IEEEtran}

\usepackage{cite}
\ifCLASSINFOpdf
   \usepackage[pdftex]{graphicx}
\else
   \usepackage[dvips]{graphicx}
\fi

\usepackage[colorlinks,linkcolor=red,anchorcolor=blue,citecolor=green,CJKbookmarks=True]{hyperref}

\usepackage{amsmath}
\usepackage{amssymb}
\usepackage{mathtools}
\usepackage{bm}
\usepackage{xfrac}

\usepackage{algorithmic}
\usepackage{titlesec}
\usepackage{enumitem}

\usepackage{array}
\usepackage{color}
\usepackage{makecell}
\usepackage{multirow}
\usepackage{tabularx}
\newcolumntype{Y}{>{\centering\arraybackslash}X}

\usepackage{url}

\newenvironment{myitemize}{%
    \begin{list}{\textbullet}{%
            \setlength{\topsep}{1pt}   \setlength{\partopsep}{0pt}
            \setlength{\parsep}{0pt}   \setlength{\itemsep}{1pt}
            \setlength{\labelsep}{3pt} \setlength{\labelwidth}{2pt}
            \setlength{\leftmargin}{5mm}}}
    {\end{list}}
\newcommand{\PreserveBackslash}[1]{\let\temp=\\#1\let\\=\temp}
\newcolumntype{C}[1]{>{\PreserveBackslash\centering}p{#1}}
\newcolumntype{R}[1]{>{\PreserveBackslash\raggedleft}p{#1}}
\newcolumntype{L}[1]{>{\PreserveBackslash\raggedright}p{#1}}

\DeclareMathOperator\dif{d\!}

\begin{document}
%
\title{Deep Variational Network Toward \\ Blind Image Restoration}
%
%
%

\author{Zongsheng Yue, Hongwei Yong, Qian Zhao,
        Lei Zhang,~\IEEEmembership{Fellow, IEEE,} \\
        Deyu Meng,~\IEEEmembership{Member, IEEE,}
        and Kwan-Yee K. Wong,~\IEEEmembership{Senior, IEEE} \\
\thanks{Z. Yue is with the School of Mathematics and Statistics, Xi'an Jiaotong University, Xi'an, China, and the Department of Computer Science, The University of Hong Kong, Hong Kong, China (e-mail: zsyue@gmail.com).}
\thanks{H. Yong and L. Zhang are with the Department of Computing, The Hong Kong Polytechnic University,
Hong Kong, China (e-mail:\{cshyong, cslzhang\}@comp.polyu.edu.hk).}
\thanks{Q. Zhao is with the School of Mathematics and Statistics and Ministry of Education Key Lab of Intelligent Networks and Network Security, Xi'an Jiaotong University, Xi'an, China (e-mail: timmy.zhaoqian@mail.xjtu.edu.cn).}
\thanks{D. Meng is with the School of Mathematics and Statistics and Ministry of Education Key Lab of Intelligent Networks and Network Security, Xi'an Jiaotong University, Xi'an, China, and also with the Macau Institute of Systems Engineering, Macau University of Science and Technology, Taipa, Macau, China (e-mail:dymeng@mail.xjtu.edu.cn).}
\thanks{K.-Y. K. Wong is with the Department of Computer Science, The University of Hong Kong, Hong Kong, China (e-mail: kykwong@cs.hku.hk).}
\thanks{D. Meng is the corresponding author.}
}

%
%

\markboth{}%
{Shell \MakeLowercase{\textit{et al.}}: A Sample Article Using IEEEtran.cls for IEEE Journals}

%


\maketitle

\begin{abstract}
    Blind image restoration (IR) is a common yet challenging problem in computer vision. Classical model-based methods and recent deep learning (DL)-based methods represent two different methodologies for this problem, each with its own merits and drawbacks. In this paper, we propose a novel blind image restoration method, aiming to integrate both the advantages of them. Specifically, we construct a general Bayesian generative model for the blind IR, which explicitly depicts the degradation process. In this proposed model, a pixel-wise non-i.i.d. Gaussian distribution is employed to fit the image noise. It is with more flexibility than the simple i.i.d. Gaussian or Laplacian distributions as adopted in most of conventional methods, so as to handle more complicated noise types contained in the image degradation. To solve the model, we design a variational inference algorithm where all the expected posteriori distributions are parameterized as deep neural networks to increase their model capability. Notably, such an inference algorithm induces a unified framework to jointly deal with the tasks of degradation estimation and image restoration. Further, the degradation information estimated in the former task is utilized to guide the latter IR process. Experiments on two typical blind IR tasks, namely image denoising and super-resolution, demonstrate that the proposed method achieves superior performance over current state-of-the-arts. The source code is available at \url{https://github.com/zsyOAOA/VIRNet}.
\end{abstract}


\section{Introduction}\label{sec:introduction}
\IEEEPARstart{I}{mage} restoration (IR) is an active research topic in the fields of signal processing and
computer vision. It aims at recovering the latent high-quality image $\bm{z}$ from the observed corrupted counterpart $\bm{y}$, i.e.,
\begin{equation}
    \bm{y}=H\bm{z}+\bm{n},
    \label{eq:image-corrupted}
\end{equation}
where $H$ is the degradation operator, and $\bm{n}$ is image noise. With
different degradation settings for $H$, Eq. (\ref{eq:image-corrupted}) represents different IR tasks. For example,
the classical IR tasks, such as image denoising, deblurring, and super-resolution, can be easily obtained by setting
$H$ as an identity matrix, a blurring operator, and a composition of blurring and downsampling operators, respectively.
The difficulties of IR tasks mainly come from $H$ and $\bm{n}$. The former inclines to cause severe information loss in some
tasks, like deblurring and super-resolution, and the latter is usually complicated due to the accumulation of noises from
multiple sources e.g., capturing instrument, camera pipeline and image transmission~\cite{tsin2001statistical}.
In blind IR tasks, we need to simultaneously solve the problems of degradation estimation and image restoration,
which makes it more challenging.

In the past decades, plenty of IR methods have been proposed under the maximum a posteriori (MAP) framework. From the Bayesian perspective, it generally involves a likelihood term and a prior term. More specifically, the likelihood term encodes the image degradation process of Eq.~\eqref{eq:image-corrupted}, while the prior term reflects our subjective knowledge of the latent high-quality image. Most of these methods mainly focused on designing more effective image priors so as to alleviate the ill-posedness of IR tasks. Commonly used image priors include total variation (TV)~\cite{rudin1992nonlinear}, non-local similarity~\cite{buades2005non,dabov2007image}, sparsity~\cite{mairal2007sparse,dong2012nonlocally,xu2018trilateral}, low-rankness~\cite{dong2012nonlocal,gu2014weighted,xu2017multi} and so on. In contrast, other works focused on the likelihood term by constructing more flexible noise distributions, e.g., mixture of Gaussian (MoG)~\cite{meng2013robust}, mixture of Exponential (MoEP)~\cite{cao2015low}, and Dirichlet Process mixture of Gaussian (DP-MoG)~\cite{zhu2016blind,yue2018hyperspectral,yue2019robust}. Even though these model-based methods are with highly intuitive physical meanings and also generalize well in most scenarios, they still have evident defects. Firstly, these methods are always time-consuming, since they require to re-solve the whole model for any new testing images. Such a one-by-one optimized paradigm tends to bring up large computational burdens, making them very hard to be applied in real applications. Secondly, limited by the manually designed likelihood and image priors, which usually cannot faithfully represent the image knowledge, they struggle to handle some complex modeling problems in real cases, such as the blind IR tasks with complicated image degradation.

Different from the aforementioned model-based methods, current deep learning (DL)-based methods represent another research trend. Their core idea is to employ the deep neural networks (DNNs), being with powerful fitting capability, to directly learn the image knowledge from a large amount of pre-collected image pairs in an end-to-end training manner. Dong~\textit{et al.}~\cite{dong2014learning} and Zhang~\textit{et al.}~\cite{zhang2017beyond} firstly proposed SRCNN and DnCNN that surpassed classical model-based methods in image super-resolution and denoising, respectively. Subsequently, many DL-based methods~\cite{mao2016image,Zhang2018,zhang2018ffdnet,Zhang2019,guo2019toward,Anwar_2019_ICCV,yue2020dual, zamir2021multi,liang2021mutual} were proposed and they achieved unprecedented successes in the field of IR. While they have achieved huge boost in performance, most of them ignore the modeling mechanism underlying the image degradation, especially the image noise. To be specific, the $L_2$ or $L_1$ loss commonly used in current DL-based methods indeed implies that the noise $\bm{n}$ in Eq.~\eqref{eq:image-corrupted} obeys the independent and identically distributed (i.i.d.) Gaussian or Laplacian distribution~\cite{meng2013robust,cao2015low}. This, however, always deviates from true noise configurations in real cases. For example, the camera sensor noises in practical image denoising are signal-dependent and spatially variant, and thus evidently non-i.i.d. in statistics. Neglecting such intrinsic noise properties will certainly harm the generalization capability of the model in real scenarios with complicated noises.

As analyzed above, the model-based methods are capable of encoding the image degradation through the likelihood, but hindered by the limited model capacity and the slow inference speed. In contrast, the DL-based methods, equipped with DNNs, are with large model capacity and powerful non-linear fitting capability. What's more, in the testing phase, these methods are much faster than the model-based method, since they only need one feedforward pass for any newly coming image. This naturally inspires us to develop a new IR method, which
is expected to combine both the advantages of the classical model-based methods and recent DL-based methods. In this work, we take one step forward along this research line by proposing a deep variational model for blind IR. It first constructs a traditional probabilistic model for IR, and then embeds the powerful DNNs into its posteriori inference to increase the model capacity. Specifically, the contributions of this work can be summarized from two aspects, namely model construction and algorithm design, as follows:

On one hand, a Bayesian generative model is built for general IR tasks, and thus naturally inherits the advantages
of classical model-based methods. Furthermore, we also consider a more complicated degradation
process when building our model in this study:
\begin{myitemize}
    \item Instead of the simple i.i.d. Gaussian or Laplacian noise assumptions in most of the current methods, a pixel-wise non-i.i.d.
        distribution is adopted in our model to handle more complicated noise types.
        In essence, such noise model induces a learnable re-weighted loss purely relying on data-self, thus is more flexible.
    \item A concise kernel prior is specifically designed for super-resolution, which makes our model able to deal with the task of blind image super-resolution.
\end{myitemize}

On the other hand, we elaborately design an amortized variational inference (VI) algorithm to solve the proposed generative model. Compared with the classical mean field VI methodology, two-fold substantial modifications are made to better comply with blind IR tasks:
\begin{myitemize}
    \item Different from the commonly used independent factorization strategy in VI, we factorize the expected posterior,
        namely the joint distribution of the degradation information and the latent clean image, into a conditional
        form. Such a formulation derives a unified framework to simultaneously deal with the tasks of degradation estimation
        and image restoration, in which the degradation information estimated by the former provides sound guidance for
        the subsequent restoration task.
    \item To largely increase the fitting capability of our model, the desirable posteriori distributions
        are parameterized by DNNs, and then optimized in an amortized manner during training. In the testing phase, the well-trained
        model is capable of fastly inferring the posteriori distribution of any new testing image in an explicit manner, and thus evidently more efficient
        than the classical model-based methods.
\end{myitemize}

In summary, this work aims to explore a novel modeling paradigm, which is expected to integrate the merits both of the classical model-based methods and recent DL-based methods, for the IR problem. A preliminary version of this work has been published in NeurIPS 2019~\cite{yue2019variational} which focuses only on image denoising. This present work makes substantial improvements on model construction, the inference algorithm, and the empirical evaluations over the conference version. Especially, we consider a more general degradation process (i.e., Eq.~\eqref{eq:image-corrupted}) to build the Bayesian generative model, making it capable of handling more complicated and general IR tasks, such as blind image super-resolution.

The remainder of the paper is organized as follows: Section 2 introduces the related work. Section 3 proposes our generative model, and discusses two typical IR tasks. Section 4 presents the designed stochastic VI algorithm for solving our model. In Section 5, experiments are demonstrated to evaluate the performance of our method. Section 6 finally concludes the paper.

\section{Related Works}
In this section, we first review model-based and DL-based IR methods. We then briefly summarize recent explorations
that attempts to combine both of these two methodologies. 

\subsection{Model-based Methods}\label{subsec:related_work_model}
Most of the classical model-based methods can be formulated into the MAP framework, which contains a likelihood (fidelity) term and a prior (regularization) term from the Bayesian perspective. Relevant developments thus mainly focused on these two terms.

\noindent\textbf{Prior Modeling Methods.} Aiming at alleviating the ill-posed issue of IR, many studies attempted to exploit rational image prior knowledge. Statistical regularities exhibited in images were firstly employed, e.g., TV denoising~\cite{rudin1992nonlinear} and wavelet coring~\cite{simoncelli1996noise}. Then, NLM~\cite{buades2005non} and BM3D~\cite{dabov2007image} were both proposed for denoising based on the non-local self-similarity prior, meaning that small image patches in a non-local image area possess similar configurations. Later, low-rankness~\cite{dong2012nonlocal,gu2014weighted,xu2017multi} and sparsity~\cite{mairal2007sparse,dabov2007image,dong2012nonlocally,xu2018trilateral} priors, which also aim to explore the characteristics of image patches, became popular and were widely used in IR tasks. To further increase the model's capacity and expression ability, some other methods moved from analytical technologies to data-driven approaches. E.g., Roth and Black~\cite{roth2009fields} proposed the fields of experts (FoE) to learn image priors. Barbu~\cite{barbu2009learning} trained a discriminative model for the Markov random field (MRF) prior, while Sun and Tappen~\cite{sun2012separable} proposed a non-local range MRF (NLR-MRF) model. More related works can be found in \cite{vemulapalli2016deep,qiao2017learning}.

\noindent\textbf{Noise Modeling Methods.} Different from the prior modeling methods, noise modeling methods focus on the likelihood (fidelity) term of the MAP framework. In fact, the widely used $L_1$ or $L_2$ loss functions implicitly make the i.i.d. Gaussian or Laplacian assumptions on image noise, which often underestimates the complexity of real noise. Based on this observation, Meng \textit{et al.} \cite{meng2013robust} proposed the MoG noise modeling method under the low-rankness framework. Furthermore, Zhu \textit{et al.}~\cite{zhu2016noise,zhu2016blind} and Yue \textit{et al.}~\cite{yue2018hyperspectral,yue2019robust} both introduced the non-parametric Dirichlet Process into MoG to increase its flexibility, leading to the adaptive adjustment for the component number of MoG.

\subsection{DL-based Methods}
DL-based methods represent a data-driven trend for the IR task. They straightforwardly train an explicit mapping function parameterized by DNN on a large amount of image pairs. The earliest convolution neural network (CNN) method can be traced back to~\cite{jain2009natural}, in which a five-layer network was employed. Some auto-encoder-based methods were then proposed in \cite{xie2012image,agostinelli2013adaptive}. Due to the insufficient research on DL technologies, however, these methods were inferior to the model-based methods in performance.

The first significant improvement of DL-based methods was achieved by \cite{burger2012image} which obtained comparable performance with BM3D~\cite{dabov2007image} in denoising task using a plain multilayer perceptron. Subsequently, Dong \textit{et al.}~\cite{dong2014learning} proposed the first CNN model for super-resolution, and it outperformed the classical model-based methods. With the advances of deep learning, Zhang \textit{et al.}~\cite{zhang2017beyond} trained a deep CNN model named DnCNN and achieved state-of-the-art performance in denoising, JPEG deblocking, and super-resolution. Since then, the DL-based methods began to dominate the research trend in almost all of the IR tasks, especially in denoising~\cite{mao2016image,zhang2018ffdnet,tai2017memnet,Anwar_2019_ICCV,zamir2021multi} and super-resolution~\cite{shi2016real, ledig2017photo,wang2018esrgan,zhang2018residual,gu2019blind,xu2020unified,zhang2021designing}.

Inspired by the development of generative adversarial network (GAN)~\cite{goodfellow2014generative}, some DL-based methods also followed the research line of noise modeling introduced in Sec.~\ref{subsec:related_work_model}. Typically, Chen \textit{et al.}~\cite{chen2018image} proposed a noise generator to simulate the real noise under the adversarial training mechanism, and Kim \textit{et al.}~\cite{kim2019grdn} further introduced some camera settings (e.g., ISO level and shutter speed) into the generator as extra guidance. More recently,
Yue \textit{et al.}~\cite{yue2020dual} proposed a dual adversarial loss to implement the noise removal and noise generation tasks into one unique Bayesian framework. Different from these GAN-based implicit noise modeling manners, this study adopts a more powerful and flexible  non-i.i.d. Gaussian distribution to explicitly model the image noise, which avoids the instability in training GAN.

\subsection{Some Relevant Explorations}
The research to combine the model-based methods and the DL-based methods has attracted increasing attention in recent years. Some
significant explorations have been attempted towards this goal.

Deep plug-and-play methods~\cite{zhang2017learning,tirer2018image,ryu2019plug,zhang2019deep,zhang2021plug} usually replace the denoising sub-problem in model-based methods with one or multiple pre-trained DNN denoisers, to leverage the abundant image prior knowledge learned by such deep denoisers. Due to the lack of end-to-end training, they always rely on tedious hyper-parameters tuning to guarantee stable performance. To alleviate this drawback, deep unfolding methods~\cite{eboli2020end,zhang2020deep,huang2020unfolding,kong2021deep,zheng2021deep} take a step forward by embedding the DNNs into traditional optimization algorithms (e.g., HQS, ADMM). Attributed to the end-to-end training manner, these deep unfolding methods achieve promising performance in some IR tasks.

Deep image prior (DIP)~\cite{ulyanov2018deep} and its related methods~\cite{gandelsman2019double,ren2020neural,liang2021flow,yue2020bsrdm} represent another significant approach along this research line. These methods aim to seek a corresponding DNN model that maps the pre-sampled noise to the desirable clean image for any given corrupted image under the MAP framework. Similar to the model-based methods, they are mainly limited by the time-consuming optimization process in the testing phase.

In this study, we develop a general and novel deep IR model which is evidently different from the above research approaches. From the model perspective, a Bayesian generative model is constructed for general IR tasks, naturally inheriting the capability of modeling image degradation from the classical model-based methods. From the algorithm perspective, we design an amortized VI algorithm that parameterizes all the involved posteriori distributions with DNNs. The embedded DNNs in our algorithm equip it with powerful fitting ability and fast testing speed as recent DL-based methods.

\section{The Proposed Method}
\subsection{Basic Settings}
In this paper, we consider two commonly used settings on the degradation operator $H$ of Eq.~\eqref{eq:image-corrupted}. In the first case, $H$ is an identity matrix, corresponding to the task of image denoising. The difficulties of this task are naturally attributed to the complexity of the image noise, which is often spatially variant and signal-dependent in real scenarios. Besides, their statistics (e.g., the noise levels) are always unknown in blind image denoising. It is thus necessary to devise methods to estimate the noise distribution and recover the latent high-quality image simultaneously.

In the second case, we consider a more general IR task, namely image super-resolution, in which $H$ is a composition of blurring and downsampling. The downsampling operation leads to serious information loss, especially in the case of large scale factors, making it more challenging compared with denoising. Similarly, blind super-resolution also involves two sub-tasks, i.e., estimating the degradation information, including both the blur kernel and noise distribution, and restoring the high-resolution image.

In addition, we briefly introduce some necessary settings on the training and testing data. The training data consists of multiple triplets, i.e., $\mathcal{D}=\{\bm{y}^{(j)}, \bm{x}^{(j)}, H^{(j)}\}_{j=1}^N$, where $\bm{y}^{(j)}$ and $\bm{x}^{(j)}$ denote the corrupted image and the underlying high-quality image, respectively. $H^{(j)}$ represents the degradation operator, which is an identity matrix for denoising and a blur kernel for super-resolution. The superscript $j$ of $H^{(j)}$ indicates that the degradation model varies from one sample to another in our training data. It should be noted that, in the real-world image denoising task, $\bm{x}^{(j)}$ is usually estimated by averaging several noisy images taken under the same camera conditions~\cite{Abdelhamed2018}. In the testing phase, given only one corrupted image, our goal is to first estimate the degradation information, and then recover the high-quality image based on such information.

Next, we formulate a rational Bayesian generative model for the tasks of denoising and super-resolution.

\subsection{Model Formulation for Denoising}\label{subsec:model_denoising}
Let $\{\bm{y}, \bm{x}\}\in \mathcal{D}$ denote any noisy/noise-free image pair in the training dataset. For the noisy image $\bm{y}$, we assume that it is generated as follows:
\begin{equation}
    y_i \sim \mathcal{N}(y_i|z_i, \sigma_i^2), ~ i=1,2, \cdots, d,
 \label{eq:noniid-generative-model}
\end{equation}
where $\mathcal{N}(\cdot|\mu, \sigma^2)$ denotes the Gaussian distribution with mean $\mu$ and variance $\sigma^2$, $\bm{z}$ denotes the latent clean image, $d$ is the number of pixels in the noisy image. Notably, different from the commonly used i.i.d. Gaussian/Laplacian assumption, we model the image noise as pixel-wise non-i.i.d. Gaussian distribution in Eq.~\eqref{eq:noniid-generative-model}. Such an non-i.i.d. noise assumption largely increases the degrees of freedom of the noise distribution and is thus expected to better fit complicated real noise as illustrated in Sec.~\ref{subsubsec:noise_est}.

Next, we introduce some prior knowledge for the latent clean image $\bm{z}$ and the noise variance map $\bm{\sigma}^2$. In the real dataset, $\bm{x}$ provides an approximate estimation to the underlying clean image $\bm{z}$ by some statistical method~\cite{Abdelhamed2018}. Therefore, it can be embedded into the following prior distribution as a constraint for $\bm{z}$:
\begin{equation}
    z_i \sim \mathcal{N}(z_i|x_i, \varepsilon_0^2), ~ i=1,2,\cdots, d,
    \label{eq:prior-z}
\end{equation}
where $\varepsilon_0$ is a hyper-parameter that reflects the closeness between $\bm{x}$ and $\bm{z}$. In some synthetic experiments where the underlying clean image is accessible, $\bm{x}$ is indeed the true clean image $\bm{z}$, which can be easily formulated by setting $\varepsilon_0$ as a small number close to $0$. Under this setting, Eq.~\eqref{eq:prior-z} degenerates to the Dirac distribution centered at $\bm{x}$.

As for the variance map $\bm{\sigma}^2$, we construct the following inverse Gamma distribution as its conjugate prior~\cite{Bishop2006}:
\begin{equation}
    \sigma_i^2 \sim \text{IG}\left(\sigma_i^2\bigg\vert\alpha_0-1, \alpha_0\xi_i\right), i=1,2,\cdots, d,
    \label{eq:prior-sigma}
\end{equation}
where
\begin{equation}
    \bm{\xi} = \mathcal{G}\left((\hat{\bm{y}}-H\hat{\bm{x}})^2;p\right).
    \label{eq:xi}
\end{equation}
$\mathcal{G}(\cdot;p)$ is the Gaussian filter with window size $p$, $\hat{\bm{y}}, \hat{\bm{x}}\in \mathbb{R}^{h\times w}$ are the matrix (image) forms of $\bm{y}, \bm{x}\in \mathbb{R}^d$,  and $\alpha_0$ is the shape\footnote{For an inverse Gamma distribution $\text{IG}(\cdot|\alpha,\beta)$, we usually call $\alpha$ and $\beta$ as the shape parameter and the scale parameter respectively.} parameter of the inverse Gamma distribution.

\begin{figure}
    \centering
    \includegraphics[scale=0.45]{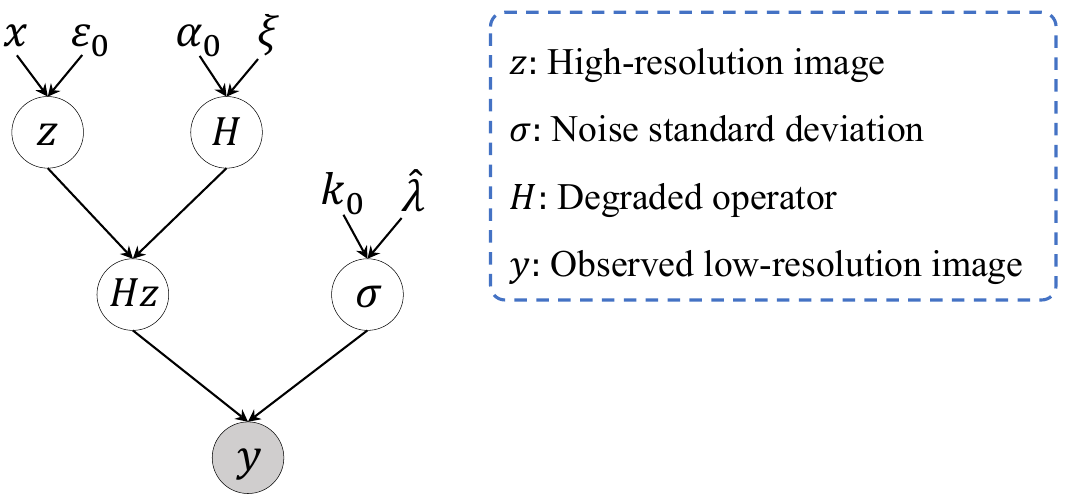}
    \vspace{-3mm}
    \caption{Illustration of the graphical model of the proposed method.}
    \label{fig:graph}
\end{figure}
Actually, $\xi_i$ in Eq.~\eqref{eq:xi} provides an estimation for the variance $\sigma_i^2$. It is calculated using a Gaussian filter in the $p\times p$ window centered at the $i$-th pixel. The elaborate design on the shape and scale parameters (i.e., $\alpha_0-1$ and $\alpha_0\xi_i$) in Eq.~\eqref{eq:prior-sigma} guarantees that the mode of this prior distribution is $\xi_i$ exactly. The hyper-parameter $\alpha_0$ controls the strength of this prior distribution, and it is set as $\sfrac{p^2}{2}$ in our method following~\cite{Yong2018}. More explanations on this hyper-parameter can be found in Appendix~\ref{subsec:IGDistribution_supp}.

\begin{figure*}[t]
    \centering
    \includegraphics[scale=0.58]{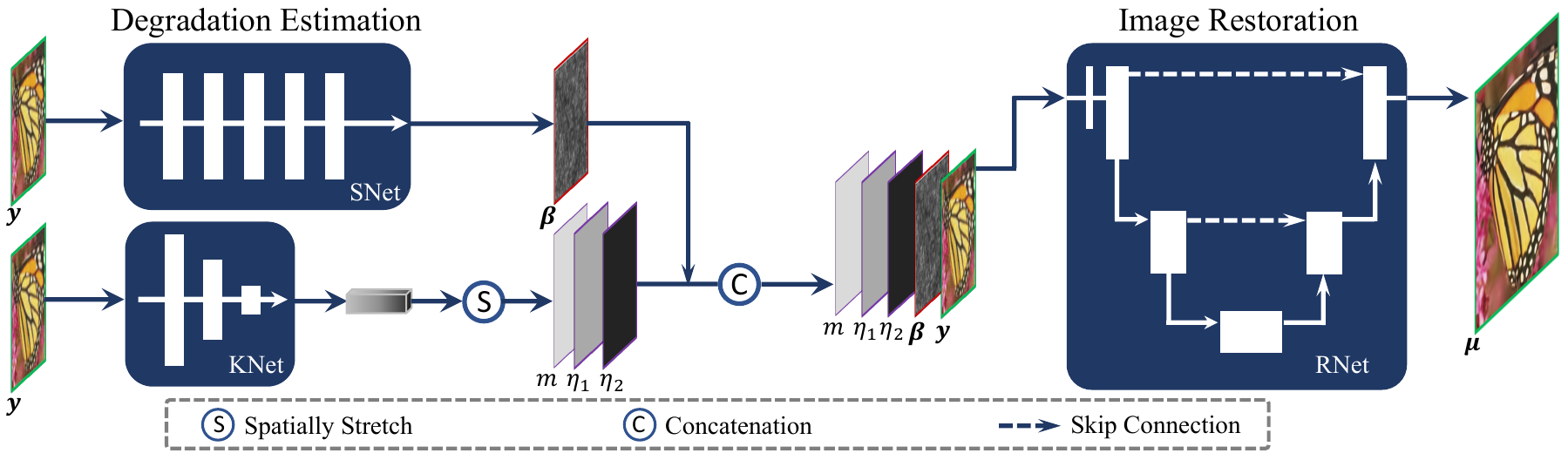}
    \vspace{-4mm}
    \caption{The inference framework of the proposed generative model for blind image super-resolution.
        It can be decomposed into two sub-tasks of degradation estimation and image restoration. Given any
        corrupted image $\bm{y}$, we firstly infer the posteriori parameters $\bm{\beta}$ of $q(\bm{\sigma}^2|\bm{y})$ by
        \textit{SNet} and $\{m,\eta_1, \eta_2\}$ of $q(\bm{\varLambda}|\bm{y})$ by \textit{KNet} in the phase of
        degradation estimation, and then recover the desirable high-quality image (i.e., the mean value $\bm{\mu}$ of
        $q(\bm{z}|\bm{y},\bm{\sigma}^2,\bm{\varLambda})$) by \textit{RNet} under the guidance of the estimated
        degradation information.}
    \label{fig:framework}
\end{figure*}
\subsection{Extension to Blind Super-resolution}
For the problem of image super-resolution, the degradation model in Eq.~\eqref{eq:image-corrupted} can be reformulated as:
\begin{equation}
    \bm{y}=(\bm{z}*\bm{k})\downarrow_s+\bm{n},
    \label{eq:SR_degradation}
\end{equation}
where $\bm{k}$ denotes the blur kernel, $*$ is the convolution operator, and $\downarrow_s$ is the downsampler with scale factor $s$. Based on this degradation model, we extend the noise assumption of Eq.~\eqref{eq:noniid-generative-model} as follows:
\begin{equation}
    y_i \sim \mathcal{N}(y_i|[(\bm{z}*\bm{k})\downarrow_s]_i, \sigma_i^2), ~ i=1,2, \cdots, d,
 \label{eq:noniid_assump_SR}
\end{equation}
where $[\bm{x}]_i$ represents the $i$-the pixel of $\bm{x}$, i.e., $x_i$.

To handle blind super-resolution, the most challenging part is how to model the blur kernel. Recently, lots of related literature~\cite{Riegler2015,zhang2018learning,huang2020unfolding,wang2021unsupervised,liang2021flow} have found that the anisotropic Gaussian kernels are sufficient to guarantee pleasing results for image super-resolution. In this study, we follow these related works and adopt the anisotropic Gaussian kernels. Thus, the blur kernel $\bm{k}$ with size $(2r+1)\times (2r+1)$ can be defined as:
\begin{align}
    k_{ij} &= g(\lambda_1^2, \lambda_2^2, \rho) \notag \\
    &=\frac{1}{2\pi \lambda_1 \lambda_2 \sqrt{1-\rho^2}} \exp\left\{ -\frac{1}{2}\bm{S}^T\bm{\Sigma}^{-1}\bm{S} \right\},
    \label{eq:kennel_define}
\end{align}
where $\bm{\Sigma}=\begin{bmatrix}
    \lambda_1^2            & \lambda_1\lambda_2\rho \\
    \lambda_1\lambda_2\rho & \lambda_2^2   \\
\end{bmatrix}$ is the covariance matrix, $\rho$ is the Pearson correlation coefficient,
$\bm{S}=\begin{bmatrix} i& j \\ \end{bmatrix}^T$ denotes the spatial coordinate with $i,j\in \{-r,\cdots,r\}$. By denoting $\bm{\varLambda}=\{\rho,\lambda_1^2, \lambda_2^2\}$, one can easily observe that the blur kernel $\bm{k}$ is completely determined by $\bm{\varLambda}$ when the kernel size is fixed. This inspires us to design a prior distribution for $\bm{\varLambda}$ instead of $\bm{k}$.

In essence, Eq.~\eqref{eq:kennel_define} represents the blur kernel $\bm{k}$ through two Gaussian distributions with variance parameters $\lambda_1^2$ and $\lambda_2^2$ along the horizontal and vertical directions, and their correlations is depicted by $\rho$. For $\{\lambda_1, \lambda_2\}$ and $\rho$, we impose the inverse Gamma~\cite{Bishop2006} and Dirac distributions for them as prior constraints, respectively, i.e.,
\begin{equation}
    \bm{\varLambda} \sim p(\bm{\varLambda})
    =\text{Dirac}(\rho|\hat{\rho})\prod_{l=1}^2 \text{IG}(\lambda_l^2|\kappa_0-1, \kappa_0*\hat{\lambda}_l^2),
    \label{eq:kernle_prior_dirac}
\end{equation}
where $\{\hat{\rho}, \hat{\lambda}_1^2, \hat{\lambda}_2^2\}$ reflect the corresponding true kernel information contained in the training data. Similar to the $\alpha_0$ of Eq.~\eqref{eq:prior-sigma}, $\kappa_0$ is also a hyper-parameter controlling the shape of the inverse Gamma distributions. For the purpose of easy optimization, we employ a Gaussian distribution with a small variance to approximate the Dirac distribution. Thus, based on Eq.~\eqref{eq:kennel_define}, the blur kernel $\bm{k}$ is modelled as:
\begin{gather}
    \bm{k} = g(\bm{\varLambda}), \label{eq:kernel_func}\\
    \bm{\varLambda} \sim p(\bm{\varLambda})
    =\mathcal{N}(\rho|\hat{\rho}, r_0^2)\prod_{l=1}^2 \text{IG}(\lambda_l^2|\kappa_0-1, \kappa_0*\hat{\lambda}_l^2),
    \label{eq:kernel_prior_gauss}
\end{gather}
where the variance $r_0^2$ is empirically set as $1e\text{-}4$ throughout all our experiments.

Combining Eqs.~\eqref{eq:prior-z}-\eqref{eq:xi}, \eqref{eq:noniid_assump_SR}, and \eqref{eq:kernel_func}-\eqref{eq:kernel_prior_gauss} (or Eqs.~\eqref{eq:noniid-generative-model}-\eqref{eq:xi}) together, a full Bayesian model for blind image super-resolution (or image denoising) can be obtained. The corresponding graphical model is depicted in Fig.~\ref{fig:graph}. Our goal then turns to infer the posterior of latent variables $\{\bm{z}, \bm{\sigma}^2, \bm{\varLambda}\}$ (or $\{\bm{z}, \bm{\sigma}^2\}$) from $\bm{y}$, namely $p(\bm{z}, \bm{\sigma}^2, \bm{\varLambda}|\bm{y})$ (or $p(\bm{z}, \bm{\sigma}^2|\bm{y})$).

\section{Stochastic Variational Inference}\label{sec:inference}
In this section, a stochastic variational inference algorithm is designed for the proposed generative model. In the following part, we take blind super-resolution problem as an example to present our algorithm, since it can be easily degenerated into the denoising task by setting the blur kernel as Dirac delta function and the scale factor as 1.

\subsection{Form of Variational Posterior}
Inspired by the VI techniques~\cite{Bishop2006}, we firstly construct a variational distribution
$q(\bm{z},\bm{\sigma}^2,\bm{\varLambda}|\bm{y})$ to approximate the true posterior
$p(\bm{z},\bm{\sigma}^2,\bm{\varLambda}|\bm{y})$ led by our generative model. The variational posteriori distribution
is then conditionally factorized as:
\begin{equation}
    q(\bm{z},\bm{\sigma}^2,\bm{\varLambda}|\bm{y}) = q(\bm{z}|\bm{y},\bm{\sigma}^2,\bm{\varLambda})
                                                     q(\bm{\sigma}^2|\bm{y})
                                                     q(\varLambda|\bm{y}).
    \label{eq:poster-factorization}
\end{equation}
Next, we begin to design specific forms for these three factorized posteriori distributions.

The conjugate prior of Eq.~\eqref{eq:prior-sigma} inspires us to assume $q(\bm{\sigma}^2|\bm{y})$ as the
following inverse Gamma distribution:
\begin{equation}
    q(\bm{\sigma}^2|\bm{y}) = \prod_i^d \text{IG}(\sigma_i^2|\alpha_0-1, \alpha_0\beta_i(\bm{y};W_S)),
    \label{eq:posterior-sigma}
\end{equation}
where $\beta_i(\bm{y};W_S)$ is a mapping function parameterized as a DNN named as sigma network (\textit{SNet}) with parameters $W_S$. It aims to predict the scale parameter of $q(\bm{\sigma}^2|\bm{y})$ directly from the corrupted image $\bm{y}$. As for the shape parameter of $q(\bm{\sigma}^2|\bm{y})$, we simply fix it as the same to the prior distribution, i.e., $\alpha_0-1$. Different from the strategy of setting them both as learnable parameters in our previous version~\cite{yue2019variational}, such a modification largely simplifies the evidence lower bound of Sec.~\ref{subsec:ELBO} and also makes our algorithm more stable during training. Similarly, we formulate $q(\bm{\varLambda}|\bm{y})$ as:
\begin{align}
     q(\bm{\varLambda}|\bm{y}) 
    &=\mathcal{N}(\rho|m(\bm{y};W_K),r_0^2) \notag \\
    &\mathrel{\phantom{=}} \prod_{l=1}^2\text{IG}(\lambda_l^2|\kappa_0-1,\kappa_0\eta_l(\bm{y};W_K)),
    \label{eq:posterir_factorize_kernel}
\end{align}
where $m(\bm{y};W_K)$ and $\eta_l(\bm{y};W_K)$ are jointly parameterized as a DNN with parameters $W_K$, named as kernel network (\textit{KNet}). It takes the low-resolution image $\bm{y}$ as input and outputs the posteriori parameters for $q(\bm{\varLambda}|\bm{y})$.

As for $q(\bm{z}|\bm{y},\bm{\sigma}^2,\bm{\varLambda})$, we set it as Gaussian distribution:
\begin{equation}
    q(\bm{z}|\bm{y},\bm{\sigma}^2,\bm{\varLambda}) = \prod_i^d
                              \mathcal{N}(z_i|\mu_i(\bm{y},\bm{\sigma}^2,\bm{\varLambda};W_R), \varepsilon_0^2),
    \label{eq:posterior-z}
\end{equation}
where $\mu_i(\bm{y},\bm{\sigma}^2,\bm{\varLambda};W_R)$ represents the mapping function to evaluate the mean value of the posteriori Gaussian distribution of $\bm{z}$. Naturally, it is parameterized as a DNN with parameters $W_R$, named as restoration network (\textit{RNet}). For the ease of training, we set the variance parameter of this posteriori distribution as a constant, i.e., $\varepsilon_0^2$, being the same with that of the prior distribution in Eq.~\eqref{eq:prior-z}.

It is necessary to emphasize that the posteriori distribution $q(\bm{z}|\bm{y},\bm{\sigma}^2, \bm{\varLambda})$ is conditioned on $\bm{\sigma}^2$ and $\bm{\varLambda}$, which means that \textit{RNet} depends on the noise variance map estimated by \textit{SNet} and the kernel information predicted by \textit{KNet}. Generally speaking, the conditional assumption of Eq.~\eqref{eq:poster-factorization} decomposes the task of blind super-resolution into two cascaded sub-tasks, namely degradation estimation implemented by \textit{SNet} and \textit{KNet}, and non-blind image restoration implemented by \textit{RNet}. The whole inference procedure is shown in Fig.~\ref{fig:framework}.

\vspace{1mm}
\noindent{\textbf{Remark}}. In Eq.~\eqref{eq:posterior-sigma}, the mode of $q(\bm{\sigma}^2|\bm{y})$
is just equal to $\bm{\beta}(\bm{y};W_S)$, which is predicted by \textit{SNet}. In other words, we leverage \textit{SNet} to only estimate the core posteriori parameter, namely the mode, instead of the whole posteriori distribution. The reasons underlying this setting are two-fold. on one hand, this strategy inclines to alleviate the learning burden of \textit{SNet} to some extent. On the other hand, we can directly utilize the output of \textit{SNet} as an estimated variance map to solve some downstream problems that depend on pre-known noise levels. Similarly, we also employ this partial learning strategy in Eq.~\eqref{eq:posterir_factorize_kernel} and Eq.~\eqref{eq:posterior-z}.

\subsection{Evidence Lower Bound} \label{subsec:ELBO}
In this part, we first induce a rational objective function, namely the evidence lower bound (ELBO), to train our model. Then, we give a further discussion to explain the relationship between our designed ELBO and the traditional mean-square error (MSE) loss.

\subsubsection{Derivation of ELBO}
For the convenience of presentation, we simply denote $\beta_i(\bm{y};W_S)$, $m(\bm{y};W_K)$, $\eta_l(\bm{y};W_K)$, and $\mu_i(\bm{y},\bm{\sigma}^2,\bm{\varLambda};W_R)$ as $\beta_i$, $m$, $\eta_l$, and $\mu_i$ respectively. Given any corrupted image $\bm{y}$, its logarithm marginal probability can be decomposed as
\begin{align}
    \log p(\bm{y}) &= \mathcal{L}(\bm{z},\bm{\sigma},\bm{\varLambda};\bm{y}) \notag \\
           &\mathrel{\phantom{=}} + KL\Big[q(\bm{z},\bm{\sigma}^2,\bm{\varLambda}|\bm{y})
                                   \Vert p(\bm{z},\bm{\sigma}^2,\bm{\varLambda}|\bm{y})\Big],
    \label{eq:marginal_likelihood}
\end{align}
where
\begin{align}
    \mathcal{L}(\bm{z},\bm{\sigma},\bm{\varLambda};\bm{y}) &= E_{q}
    \Big[\log p(\bm{y}|\bm{z},\bm{\sigma}^2,\bm{\varLambda})p(\bm{z})p(\bm{\sigma}^2)p(\bm{\varLambda})  \notag \\
    &\mathrel{\phantom{=}} - \log q(\bm{z},\bm{\sigma}^2,\bm{\varLambda}|\bm{y}) \Big].
    \label{eq:ELBO-V1}
\end{align}
Here $E_{q}[\cdot]$ denotes the expectation w.r.t. the posteriori distribution
$q(\bm{z},\bm{\sigma}^2,\bm{\varLambda}|\bm{y})$. The second term of Eq.~\eqref{eq:marginal_likelihood} represents the KL divergence between the variational posterior $q(\bm{z},\bm{\sigma}^2,\bm{\varLambda}|\bm{y})$ and the true posterior $p(\bm{z},\bm{\sigma}^2,\bm{\varLambda}|\bm{y})$. Due to the non-negativeness of KL divergence, $\mathcal{L}(\bm{z},\bm{\sigma},\bm{\varLambda};\bm{y})$ constitutes a lower bound of $\log p(\bm{y})$, thus called ELBO. Therefore, we can naturally approximate the true posteriori $p(\bm{z},\bm{\sigma}^2,\bm{\varLambda}|\bm{y})$ with $q(\bm{z},\bm{\sigma}^2,\bm{\varLambda}|\bm{y})$ via maximizing ELBO.

Combining the factorized assumption of Eq.~\eqref{eq:poster-factorization}, the ELBO can be rewritten as
\begin{align}
    \mathcal{L}(\bm{z},\bm{\sigma},\bm{\varLambda};\bm{y})
                &= E_{q(\bm{z},\bm{\sigma},\bm{\varLambda}|\bm{y})}\Big[\log p(\bm{y}|\bm{z},\bm{\sigma}^2,\bm{\varLambda})\Big] \notag \\
                &- E_{q(\bm{\sigma},\bm{\varLambda}|\bm{y})}\Big[
                    KL\left[q(\bm{z}|\bm{y},\bm{\sigma}^2,\bm{\varLambda}) \Vert p(\bm{z})\right]
                    \Big]\notag \\
                &- KL\Big[q(\bm{\sigma}^2|\bm{y}) \Vert p(\bm{\sigma}^2)\Big] \notag \\
                &- KL\Big[q(\bm{\varLambda}|\bm{y}) \Vert p(\bm{\varLambda})\Big],
    \label{eq:ELBO_KL}
\end{align}
where $q(\bm{\sigma},\bm{\varLambda}|\bm{y})=q(\bm{\sigma}|\bm{y})q(\bm{\varLambda}|\bm{y})$.

Next, we consider how to calculate each term in Eq.~\eqref{eq:ELBO_KL} step by step.
The first term is intractable, mainly because the posterior $q(\bm{z},\bm{\sigma},\bm{\varLambda}|\bm{y})$ is parameterized as complicated forms of DNNs. Fortunately, we can use the reparameterization trick~\cite{Kingma2014} to obtain multiple differentiable samples from the posteriors, and then use them to estimate these two terms by Monte Carlo (MC) like VAE~\cite{Kingma2014}. Concretely, the re-sampling process from $q(\bm{z}|\bm{y},\bm{\sigma}^2,\bm{\varLambda})$ can be easily implemented as
\begin{equation}
    \bm{\tilde{z}} = \bm{\mu} + \varepsilon_0 \bm{\tau}, ~ \bm{\tau} \sim \mathcal{N}(\tau|\bm{0}, I).
    \label{eq:resample_z}
\end{equation}
In the training phase, we sample from $q(\bm{\varLambda}|\bm{y})$ and $q(\bm{\sigma}^2|\bm{y})$ utilizing the pathwise derivative technology~\cite{jankowiak2018pathwise}, and denote the re-sampled data example as $\tilde{\bm{\varLambda}}$ and $\tilde{\bm{\sigma}}^2$. Based on $\tilde{\bm{z}}$, $\tilde{\bm{\varLambda}}$ and $\tilde{\bm{\sigma}}^2$, the first term of Eq.~\eqref{eq:ELBO_KL} can be approximated as follows:
\begin{align}
    &\mathrel{\phantom{\approx}}
    E_{q(\bm{z},\bm{\sigma},\bm{\varLambda}|\bm{y})}\left[\log p(\bm{y}|\bm{z},\bm{\sigma}^2,\bm{\varLambda})\right] \notag \\
    &\approx -\frac{1}{2}\sum_{i=1}^d
    \Big \{\log\tilde{\sigma}_i + w_i\left(y_i - [(\tilde{\bm{z}}*\tilde{\bm{k}})\downarrow_s]_i\right)^2 \Big\},
    \label{eq:likelihood_general} 
\end{align}
where $w_i=\frac{1}{\tilde{\sigma}_i^2}$, $\tilde{\bm{k}}=g(\tilde{\bm{\varLambda}})$,
and $g(\cdot)$ is defined in Eq.~\eqref{eq:kernel_func}. Note that we have omitted a constant that is independent of the learnable parameters in Eq.~\eqref{eq:likelihood_general}.

As for the last three terms of Eq.~\eqref{eq:ELBO_KL}, they can all be analytically calculated as follows:
\begin{small}
\begin{equation}
    E_{q(\bm{\sigma},\bm{\varLambda}|\bm{y})}\Big[
        KL\left[q(\bm{z}|\bm{y},\bm{\sigma}^2,\bm{\varLambda}) \Vert p(\bm{z})\right]
        \Big] 
        = \sum_{i=1}^d \frac{(\mu_i-x_i)^2}{2\varepsilon_0^2},
        \label{eq:KL-z}  
\end{equation}
\begin{equation}
    KL\Big[q(\bm{\sigma}^2|\bm{y})\Vert p(\bm{\sigma}^2) \Big] = \sum_{i=1}^d
    \alpha_0\left(\frac{\xi_i}{\beta_i} + \log \frac{\beta_i}{\xi_i} - 1 \right), \label{eq:KL_sigma}
\end{equation}
\begin{equation}
    KL\Big[q(\bm{\varLambda}|\bm{y})\Vert p(\bm{\varLambda}) \Big]
    = \frac{(m-\hat{\rho})^2}{2r_0^2} + \sum_{l=1}^2\kappa_0
    \left(\frac{\hat{\lambda}_l}{\eta_l} + \log \frac{\eta_l}{\hat{\lambda}_i} - 1 \right).
    \label{eq:KL_kernel}
\end{equation}

Finally, we can get the expected objective function, namely the negative ELBO on the entire training dataset, to optimize the network parameters of $W_S$, $W_K$, and $W_R$ as follows:
\begin{equation}
    \min_{W_S, W_K, W_R} -\sum_{j=1}^N \mathcal{L}(\bm{z}^{(j)},\bm{\sigma}^{(j)},\bm{\varLambda}^{(j)};\bm{y}^{(j)}),
    \label{eq:loss_all}
\end{equation}
\end{small}
where $\bm{z}^{(j)}$, $\bm{\sigma}^{(j)}$, and $\bm{\varLambda}^{(j)}$ denote the posteriori parameters for the $j$-th image pair in training dataset $\mathcal{D}$.

In the testing phase, we first predict the variance parameter $\bm{\beta}$ as well as the kernel parameters $\{m, \eta_1, \eta_2\}$ using \textit{SNet} and \textit{KNet}. Subsequently, these predicted parameters are concatenated with the low-resolution image $\bm{y}$ and serve as the input of \textit{RNet}. The output of \textit{RNet}, i.e., the mean parameter $\bm{\mu}$ of the posterior of the underlying high-quality image, can be reasonably regarded as our desirable restored result. The whole inference pipeline is illustrated in Fig.~\ref{fig:framework}.

\subsubsection{Discussion} \label{subsucsec:discussion-elbo}
Let us delve further into the ELBO loss of Eq.~\eqref{eq:ELBO_KL}. The first term of this formulation presents the log-likelihood of the observed low-resolution image, enforcing that the recovered high-resolution image can be mapped back to the low-resolution one via the estimated degradation model. The other three terms correspond to the constraints for the estimated high-resolution image $\bm{z}$, the noise variance $\bm{\sigma}^2$, and the kernel parameters $\bm{\varLambda}$, respectively, as shown in Eqs.~\eqref{eq:KL-z}-\eqref{eq:KL_kernel}. Particularly noteworthy is the constraint imposed on $\bm{z}$ in Eq.~\eqref{eq:KL-z}, which takes the form of an MSE endowed with a weight factor of $\frac{1}{2\varepsilon_0^2}$. In other words, our designed ELBO loss degenerates into the commonly used MSE by setting the hyper-parameter $\varepsilon_0$ to a small value close to zero. A more thorough experimental comparison with the MSE baseline can be found in Sec.~\ref{subsec:ablation-exp}.

We now specifically consider the log-likelihood term, commonly called as data fidelity within the field of IR. Most of the existing IR methods~\cite{gu2014weighted,dong2012nonlocal} usually posit uniform importance to each element in this term, i.e., $\sum_i \left(y_i - [(\tilde{\bm{z}}*\tilde{k})\downarrow_s]_i\right)^p$. This study introduces an innovative approach whereby we exploit an adaptive manner to re-weight the data fidelity in terms of $l_2$-norm, i.e., $\sum_i w_i\left(y_i - [(\tilde{\bm{z}}*\tilde{k})\downarrow_s]_i\right)^2$, as detailed in Eq.~\eqref{eq:likelihood_general}. Each pixel is assigned with a weight of $w_i=\frac{1}{\tilde{\sigma}_i^2}$, in which $\tilde{\sigma}_i^2$ is sampled from the noise distribution estimated by \textit{SNet}. This re-weighting strategy, based on noise variance, aligns with the principles of Bayesian statistics, akin to prior literature such as~\cite{meng2013robust,zhu2016blind}. 

\begin{table*}[t]
    \centering
    \caption{The PSNR and SSIM results of different methods on three groups of testing datasets. The best
    and second best results are highlighted in \textbf{bold} and \underline{underline}, respectively. Note that * denotes  non-blind
    methods that rely on the pre-given noise-level.}
    \label{tab:psnr_noniid}
    \vspace{-4mm}
    \begin{tabular}{@{}C{1.2cm}@{}|@{}C{1.5cm}@{}|@{}C{1.2cm}@{}|
                    @{}C{1.6cm}@{} @{}C{1.8cm}@{} @{}C{1.9cm}@{} @{}C{2.0cm}@{}
                    @{}C{1.6cm}@{} @{}C{1.6cm}@{} @{}C{2.0cm}@{} @{}C{1.6cm}@{}}
        \Xhline{0.8pt}
        \multirow{2}*{Cases}&\multirow{2}*{Datasets} &\multirow{2}*{Metrics} & \multicolumn{8}{c}{Methods} \\
        \Xcline{4-11}{0.4pt}
        &         &         &NLM~\cite{buades2005non}                 &CBM3D~\cite{dabov2007image}   &DnCNN~\cite{zhang2017beyond}
                            &FFDNet$^{*}$~\cite{zhang2018ffdnet}      &S2S~\cite{quan2020self2self}
                            &Ne2Ne~\cite{huang2021neighbor2neighbor}
                            &DRUNet$^{*}$~\cite{zhang2017learning}    &VIRNet  \\
        \Xhline{0.4pt}
        \multirow{4}*{Case 1} &\multirow{2}*{CBSD68}
                    &PSNR     & 24.06   &26.73    &28.74   &28.79    &28.23   &27.92   &\underline{29.05}   &\textbf{29.28} \\
          &         &SSIM     & 0.6190  &0.7660   &0.8181  &0.8181   &0.7968  &0.7948  &\underline{0.8349}  &\textbf{0.8353}   \\
        \Xcline{2-11}{0.4pt}  &\multirow{2}*{McMaster}
                    &PSNR     & 25.08   &27.47    &29.49   &30.17    &29.87   &29.63   &\underline{30.86}   &\textbf{31.00}   \\
          &         &SSIM     & 0.6910  &0.7800   &0.8218  &0.8394   &0.8374  &0.8263  &\underline{0.8640}  &\textbf{0.8642}  \\
        \Xhline{0.4pt}
        \multirow{4}*{Case 2} &\multirow{2}*{CBSD68}
                    &PSNR     & 22.40   &25.42    &28.15   &28.42    &27.87   &25.26   &\underline{28.64}   &\textbf{28.93} \\
          &         &SSIM     & 0.5582  &0.7040   &0.7989  &0.8079   &0.7859  &0.6870  &\underline{0.8251}  &\textbf{0.8269}\\
        \Xcline{2-11}{0.4pt}  &\multirow{2}*{McMaster}
                    &PSNR     & 23.26   &25.82    &28.84   &29.74    &29.43   &27.23   &\underline{30.38}   &\textbf{30.58}   \\
          &         &SSIM     & 0.6126  &0.7120   &0.7994  &0.8315   &0.8284  &0.7314  &\underline{0.8545}  &\textbf{0.8572}   \\
        \Xhline{0.4pt}
        \multirow{4}*{Case 3} &\multirow{2}*{CBSD68}
                    &PSNR     & 24.07   &26.85    &28.64   &28.68    &28.14   &27.33   &\underline{29.12}   &\textbf{29.19}  \\
          &         &SSIM     & 0.6153  &0.7360   &0.8143  &0.8141   &0.7931  &0.7629  &\underline{0.8321}  &\textbf{0.8323}  \\
        \Xcline{2-11}{0.4pt}  &\multirow{2}*{McMaster}
                    &PSNR     & 25.13   &27.62    &29.36   &30.02    &29.75   &28.97   &\underline{30.82}   &\textbf{30.85}  \\
          &         &SSIM     & 0.6845  &0.7520   &0.8184  &0.8365   &0.8354  &0.7982  &\underline{0.8610}  &\textbf{0.8612}   \\
        \Xhline{0.8pt}
    \end{tabular}
    \vspace{-3mm}
\end{table*}

\subsection{Network Structure and Learning}
As shown in Fig.~\ref{fig:framework}, \textit{SNet} takes the corrupted image $\bm{y}$ as input and outputs the scale-related parameter of $q(\bm{\sigma}^2|\bm{y})$, achieving the goal of noise estimation. In practice, it consists of five convolution layers, and each is followed with a Leaky ReLU activation except for the first and last layers. As for the \textit{KNet}, it is designed to predict the posteriori distribution of the kernel parameter $\bm{\varLambda}$ from the corrupted image $\bm{y}$. In the implementation, we firstly employ one convolutional layer and eight channel attention blocks (CAB)~\cite{Anwar_2019_ICCV} to extract abundant feature maps, and then fuse them by one convolutional layer followed by an average pooling layer to obtain the posteriori parameters in $q(\bm{\varLambda}|\bm{y})$.

The design of \textit{RNet}, aiming to infer the conditional posteriori distribution of the desirable high-quality image, plays the most important role in blind IR. We adopt the commonly used ResUNet~\cite{zhang2020deep,zhang2021plug} in low-level vision as our backbone. It replaces the plain convolution layer in UNet~\cite{ronneberger2015u} with residual block~\cite{he2016deep}, and thus makes the gradient flow propagate much faster. Furthermore, in purpose of leveraging the estimated noise and kernel information by \textit{SNet} and \textit{KNet}, we concatenate their outputs with the corrupted image $\bm{y}$ together (see Fig.~\ref{fig:framework}), and then feed them into $\textit{RNet}$ to recover the high-resolution image. We empirically find that such a simple concatenated operation performs very well and stably in our inference framework.

It should be noted that this work does not aim to design more effective network architectures to surpass current SotA methods but to devise a probabilistic framework based on the deep variational inference for blind IR. Therefore, we simply employ the commonly used networks in low-level vision as our backbones. To better verify the generality of the proposed framework, we also test its effectiveness on other popular network architectures in Sec.~\ref{subsubsec:network-ablation}. 

\section{Experimental Results}
In this section, we evaluate the effectiveness of our proposed method on two typical IR tasks, namely image denoising and image super-resolution. We denote our \textbf{V}ariational \textbf{I}mage \textbf{R}estoration \textbf{Net}work as VIRNet for notation convenience in the following presentation.

To optimize the network, we adopted the Adam~\cite{Kingma2015} algorithm with a mini-batch size of 16 and other default settings of PyTorch~\cite{paszke2019pytorch}. The initial learning rate was set as $10^{-4}$ and decayed gradually using the cosine annealing strategy~\cite{eltit}. For computational stability, the gradient clipping strategy was also used during training. In the task of image denoising, we cropped small image patches with a size of $128\times 128$ for training. The hyper-parameter $\varepsilon_0^2$ of Eq.~\eqref{eq:prior-z} was set to be $10^{-6}$, and the window size $p$ of Eq.~\eqref{eq:xi} was set to be $7$. In the task of image super-resolution, the patch size during training was fixed as $96$, $144$, and $192$ for scale factors 2, 3, and 4 respectively. The hyper-parameter $\varepsilon_0^2$ was set to be $10^{-5}$, while the window size $p$ was set as a larger value $11$ than that in denoising, since image noise in super-resolution is usually assumed to be i.i.d. Gaussian. As for the shape parameter $\kappa_0$ of the kernel prior distribution in Eq.~\eqref{eq:kernel_prior_gauss}, we empirically set it as 50.

\subsection{Image Denoising Experiments} \label{subsec:denoising-experiment}
\subsubsection{Synthetic Non-I.I.D. Gaussian Noise Removal} \label{subsubsec:synthetica-denoising}
\begin{figure}[t]
    \centering
    \includegraphics[scale=0.38]{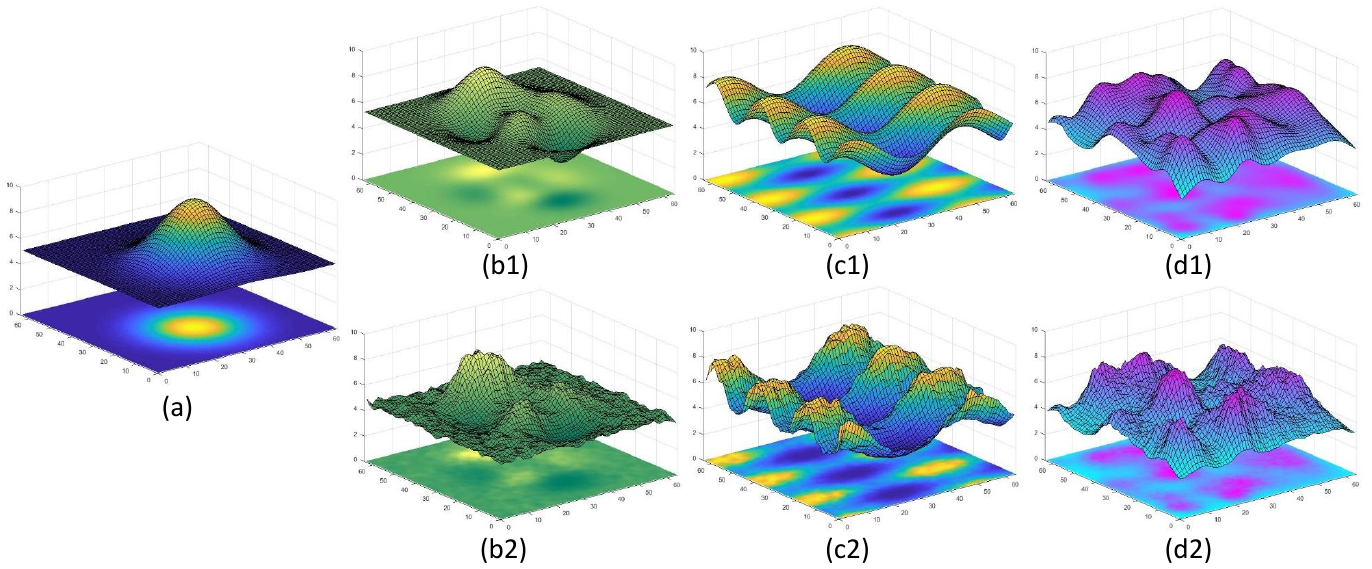}
    \vspace{-8mm}
    \caption{(a) The spatially variant map $\bm{M}$ for noise generation in training data. (b1)-(d1): Three different
    $\bm{M}$s on testing data in Cases 1-3. (b2)-(d2): Predicted $\bm{M}$s by our method on testing data.}
    \label{fig:sigmaMap}
    \vspace{-3mm}
\end{figure}

To verify the effectiveness and robustness of VIRNet under non-i.i.d. noise configurations, we synthesized a large set
of noisy/clean image pairs as training data. Similar to~\cite{zhang2018ffdnet}, a set of high-quality source images
was firstly collected as clean ones, including 432 images from BSD500~\cite{arbelaez2010contour}, 400 images from the
ImageNet~\cite{deng2009imagenet} validation set and 4744 images from Waterloo Database~\cite{Ma2017}. We then randomly
generated non-i.i.d. Gaussian noise as
\begin{equation}
    \bm{n} = \bm{n}^1 \odot \bm{M}, ~ \bm{n}^1 \sim \mathcal{N}(\bm{n}^1|\bm{0}, \bm{I}),
    \label{eq:non-iid-noise}
\end{equation}
where $\bm{I}$ is the identity matrix, and $\bm{M}$ is a spatially variant map with the same size as the source image. Finally,
the noisy image was obtained by adding the generated noise $\bm{n}$ to the source image. As for the testing images, two
commonly-used datasets, i.e., BSD68~\cite{arbelaez2010contour} and McMaster~\cite{Zhang2011}, were adopted to evaluate
the performance of different methods. Note that we totally generated four kinds of $\bm{M}$s as shown in
Fig.~\ref{fig:sigmaMap}. The first one (Fig.~\ref{fig:sigmaMap} (a)) was used for generating noisy images in training data, and
the others (Fig.~\ref{fig:sigmaMap} (b1)-(d1)) for three groups of testing data (denoted as Cases 1-3).
Under these settings, the noise in training data and testing data are evidently different, which is suitable to verify
the generalization capability of VIRNet.
\begin{figure*}[t]
    \centering
    \includegraphics[scale=1.01]{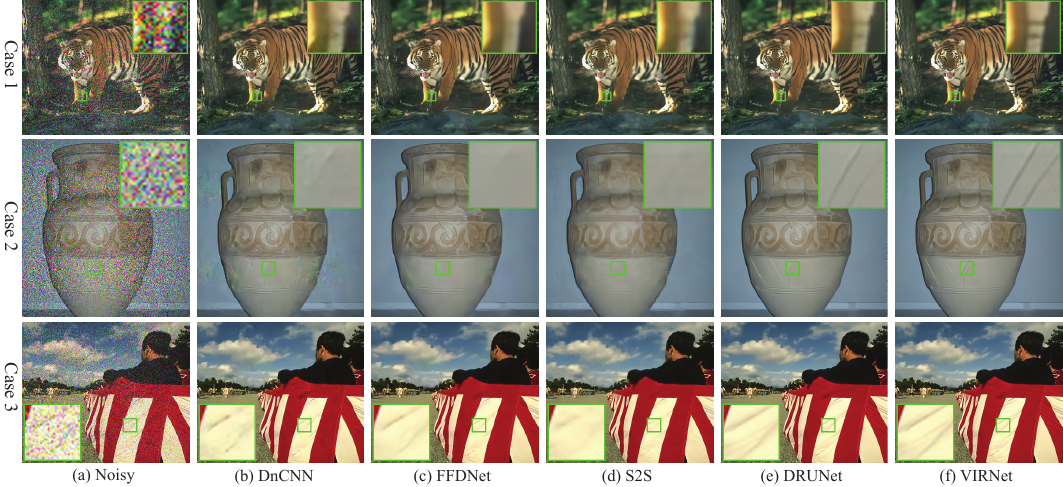}
    \vspace{-7mm}
    \caption{Denoising results of different competing methods on three typical test examples of synthetic non-i.i.d. Gaussian Noise Removal.}
    \label{fig:DN_syn}
    \vspace{-2mm}
\end{figure*}
\begin{table*}[t]
    \centering
    \caption{The PSNR/SSIM results of different methods under AWGN noises. The best and second best results
    are highlighted in \textbf{bold} and \underline{underline}, respectively. Note that * denotes non-blind methods that rely
    on the pre-given noise level.}
    \label{tab:psnr_iidawgn}
    \vspace{-4mm}
    \begin{tabular}{@{}C{1.2cm}@{}|@{}C{1.5cm}@{}|@{}C{1.2cm}@{}|
                    @{}C{1.8cm}@{} @{}C{1.8cm}@{} @{}C{2.0cm}@{} @{}C{1.5cm}@{}
                    @{}C{1.6cm}@{} @{}C{2.0cm}@{} @{}C{2.0cm}@{} @{}C{1.5cm}@{}}
        \Xhline{0.8pt}
        \multirow{2}*{Cases}&\multirow{2}*{Datasets} &\multirow{2}*{Metrics} & \multicolumn{8}{c}{Methods} \\
        \Xcline{4-11}{0.4pt}
        &         &         & CBM3D~\cite{dabov2007image}                & DnCNN~\cite{zhang2017beyond}
                            & FFDNet$^{*}$~\cite{zhang2018ffdnet}        & S2S~\cite{quan2020self2self}
                            & Ne2Ne~\cite{huang2021neighbor2neighbor}    & RNAN$^{*}$~\cite{Zhang2019}
                            & DRUNet$^{*}$~\cite{zhang2017learning}      & VIRNet  \\
        \Xhline{0.4pt}
        \multirow{4}*{$\sigma=15$} &\multirow{2}*{CBSD68}
                    &PSNR     & 33.56   &33.88    &33.87   &32.16    &33.55   &-                &\underline{34.18}   &\textbf{34.27} \\
          &         &SSIM     & 0.9237  &0.9288   &0.9288  &0.9026   &0.9262  &-                &\textbf{0.9341}     &\underline{0.9340}   \\
        \Xcline{2-11}{0.4pt}  &\multirow{2}*{McMaster}
                    &PSNR     & 34.05   &33.45    &34.65   &33.20    &34.49   &-                &\textbf{35.33}      &\textbf{35.33}   \\
          &         &SSIM     & 0.9105  &0.9034   &0.9214  &0.9014   &0.9208  &-                &\textbf{0.9323}     &\underline{0.9312}  \\
        \Xhline{0.4pt}
        \multirow{4}*{$\sigma=25$} &\multirow{2}*{CBSD68}
                    &PSNR     & 30.81   &31.22    &31.20   &30.22    &31.07   &-                &\underline{31.50}   &\textbf{31.65} \\
          &         &SSIM     & 0.8700  &0.8824   &0.8817  &0.8563   &0.8815  &-                &\underline{0.8917}  &\textbf{0.8918}\\
        \Xcline{2-11}{0.4pt}  &\multirow{2}*{McMaster}
                    &PSNR     & 31.68   &31.50    &32.34   &31.56    &32.27   &-                &\underline{33.02}   &\textbf{33.08}   \\
          &         &SSIM     & 0.8700  &0.8691   &0.8857  &0.8745   &0.8876  &-                &\textbf{0.9028}     &\underline{0.9017}   \\
        \Xhline{0.4pt}
        \multirow{4}*{$\sigma=50$} &\multirow{2}*{CBSD68}
                    &PSNR     & 27.47   &27.91    &27.95   &27.58    &27.83   &\underline{28.25}&28.15               &\textbf{28.45}  \\
          &         &SSIM     & 0.7680  &0.7885   &0.7882  &0.7716   &0.7829  &0.8010           &\underline{0.8082}  &\textbf{0.8092}  \\
        \Xcline{2-11}{0.4pt}  &\multirow{2}*{McMaster}
                    &PSNR     & 28.53   &28.61    &29.17   &29.11    &29.10   &29.69            &\underline{29.84}   &\textbf{30.02}  \\
          &         &SSIM     & 0.7894  &0.7984   &0.8138  &0.8191   &0.8147  &0.8326           &\textbf{0.8445}     &\underline{0.8433}   \\
        \Xhline{0.8pt}
    \end{tabular}
    \vspace{-3mm}
\end{table*}
\begin{table}[t]
    \centering
    \caption{Comparison results with current DL-based methods on model parameters (in M) and FLOPs (in G).}
    \label{tab:model_profile_syn}
    \vspace{-4mm}
    \begin{tabular}{@{}C{1.5cm}@{}|@{}C{1.4cm}@{}@{}C{1.5cm}@{} @{}C{1.4cm}@{} @{}C{1.5cm}@{} @{}C{1.5cm}@{}}
        \Xhline{0.8pt}
        \multirow{2}*{Metrics} & \multicolumn{5}{c}{Methods} \\
        \Xcline{2-6}{0.4pt} & DnCNN    & FFDNet  & RNAN        & DRUNet      & VIRNet \\
        \hline
        \# Param            & 0.67     & 0.85    & 8.96        & 32.64       & 10.54  \\
        \hline
           FLOPs            &175       & 56      & 3420        & 574         & 680    \\
        \Xhline{0.8pt}
    \end{tabular}
    \vspace{-3mm}
\end{table}

\vspace{1mm}
\noindent\textbf{Comparison with the SotAs}. We compare VIRNet with several current denoising methods, including two typical model-based methods NLM~\cite{buades2005non} and CBM3D~\cite{dabov2007image}, two deep self-supervised methods S2S~\cite{quan2020self2self} and Ne2Ne~\cite{huang2021neighbor2neighbor}, three supervised learning-based methods, namely DnCNN~\cite{zhang2017beyond}, FFDNet~\cite{zhang2018ffdnet}, and DRUNet~\cite{zhang2021plug}. The PSNR and SSIM results of all comparison methods on three groups of testing data are listed in Table~\ref{tab:psnr_noniid}. We can easily see that: 1) the proposed VIRNet outperforms the other methods in all cases, indicating its superiority in handling these complicated non-i.i.d. noise types; 2) on the whole, DL-based methods (including the self-supervised methods) evidently surpass classical model-based methods NLM and CBM3D, owning to the powerful non-linear fitting capability of DNNs; 3) FFDNet and DRUNet are both non-blind methods that rely on the pre-given noise level as input. In contrast, VIRNet is designed toward blind IR, and is thus able to simultaneously infer the noise distribution and remove the noise. Even so, VIRNet still achieves obvious performance improvements compared with FFDNet and DRUNet. This indicates the effectiveness of the Bayesian generative model and the variational inference framework.

Fig.~\ref{fig:DN_syn} shows the visual results of different methods under testing cases 1-3 of Table~\ref{tab:psnr_noniid}. Note that we only display the best five DL-based methods due to page limitation. It can be easily seen that the comparison methods can remove most of the noises, but also often generate over-smooth and blurry recovery, especially in the heavy-noise areas. This can be explained by the fact that they do not consider spatial noise variations. To handle such non-i.i.d. noise, the proposed VIRNet elaborately considers the noise configurations and is thus capable of preserving more image details (e.g., edges, structures) than other methods.
\begin{figure*}[t]
    \centering
    \includegraphics[scale=0.90]{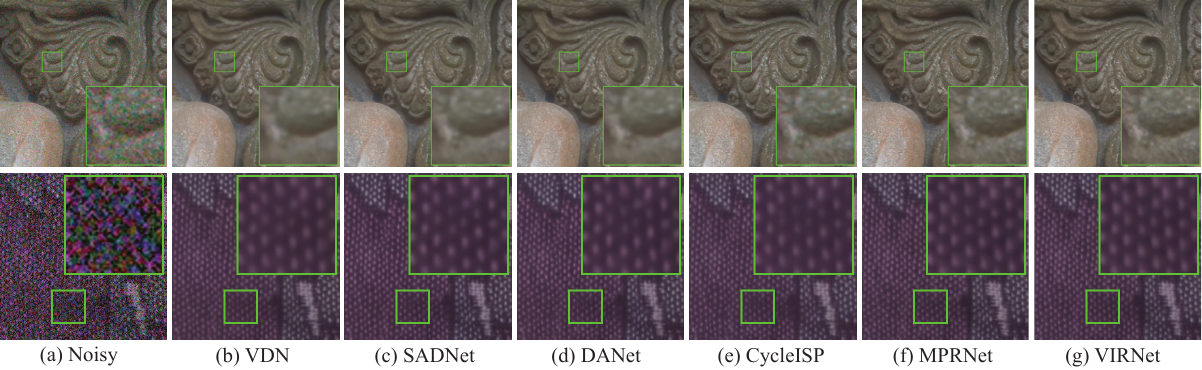}
    \vspace{-7mm}
    \caption{Denoising results of all competing methods on two typical real-world  examples from DND~\cite{Plotz2017} (upper) and SIDD~\cite{Abdelhamed2018} (lower) datasets.}
    \label{fig:DN_real}
    \vspace{-3mm}
\end{figure*}

Even though our VIRNet is designed and trained on the non-i.i.d. noise settings, it also performs well in additive white Gaussian noise (AWGN) removal
tasks. Table~\ref{tab:psnr_iidawgn} lists the average PSNR and SSIM results of different methods under three noise levels (i.e., $\sigma$=15, 25, 50) of AWGN.
In this part, we further add the method RNAN~\cite{Zhang2019} for a more thorough evaluation. It  is noteworthy that RNAN is separately trained on some
specific noise levels for AWGN, and hence we can only compare with it on the noise level 50.
It is easy to see that VIRNet obtains the best (8 out of 12 cases) or at least second-best (4 out of 12 cases) performance compared with these comparison methods.
Combining the results in Table~\ref{tab:psnr_noniid} and Table~\ref{tab:psnr_iidawgn}, it should be rational to say that the proposed VIRNet is more robust.
Specifically, it is hopeful to handle a wider range of noise types, due to its more flexible noise modeling essence.

In Table~\ref{tab:model_profile_syn}, we further list the comparison results on the number of model parameters and FLOPs with four DL-based methods.
The FLOPs listed in this table are calculated on images with a size of $512 \times 512$. It should be noted that, for the sake of fair comparisons,
the self-supervised methods S2S~\cite{quan2020self2self} and Ne2Ne~\cite{huang2021neighbor2neighbor} are not reported in Table~\ref{tab:model_profile_syn}.
It can be easily observed that VIRNet exhibits a better compromise
over current SotA methods RNAN~\cite{Zhang2019} and DRUNet~\cite{zhang2021designing}
when taking both the model parameters and FLOPs into consideration. The proposed VIRNet is thus expected with better practical applicability in real scenarios.

\subsubsection{Real-world Noise Removal}\label{subsubsec:real-denoising}
In this part, we evaluate the performance of VIRNet on two widely used real-world benchmark datasets, namely DND~\cite{Plotz2017} and SIDD~\cite{Abdelhamed2018}. DND\footnote{\url{https://noise.visinf.tu-darmstadt.de}} consists of 50 high-resolution images with realistic noise from 50 scenes taken by 4 consumer cameras, but it does not provide any other noisy/clean image pairs as training data. SIDD\footnote{\url{https://www.eecs.yorku.ca/~kamel/sidd/benchmark.php}} is another real-world denoising benchmark, containing about 30,000 real noisy images captured by 5 cameras under 10 scenes. Different from DND, each noisy image in SIDD comes with an almost noise-free counterpart as ground truth, which is estimated by some statistical methods~\cite{Abdelhamed2018}. Further, SIDD also provides a small version dataset containing 320 image pairs, called SIDD-Medium, which is commonly used as training data in recent works~\cite{yue2019variational,Anwar_2019_ICCV,yue2020dual}. To compare with them fairly, we also train VIRNet only based on the SIDD-Medium dataset. As for the metrics, we adopt PSNR and SSIM~\cite{Wang2004} calculated on the sRGB space to quantitatively evaluate different methods.
\begin{table}[t]
    \centering
    \caption{Comparisons of different methods in terms of denoising performance, the number of model parameters (in M), FLOPs (in G),
        and running time (in seconds). Note that PNGAN follows the same network architecture and the training
        strategy with MPRNet, and further improves the performance by simulating more training data. Thus we mark
        its results with \textcolor[gray]{0.5}{gray} color to denote unfair comparisons.}
    \label{tab:DN_real}
    \vspace{-4mm}
    \scriptsize
    \begin{tabular}{@{}C{1.8cm}@{}|@{}C{0.9cm}@{}@{}C{0.9cm}@{}|
                                   @{}C{0.9cm}@{}@{}C{0.9cm}@{}|
                                   @{}C{1.2cm}@{}@{}C{1.1cm}@{}@{}C{1.1cm}@{}}
        \Xhline{0.8pt}
        \multirow{2}*{Methods}            & \multicolumn{2}{c|}{SIDD}  & \multicolumn{2}{c|}{DND} & \multicolumn{3}{c}{Model Profile} \\
        \Xcline{2-8}{0.4pt}
                                          & PSNR  & SSIM  & PSNR  & SSIM  & \# Param  & FLOPs & Time   \\
        \hline
        CBM3D~\cite{dabov2007image}       & 25.65 & 0.685 & 34.51 & 0.851 & -         & -     & 21.49   \\
        DnCNN~\cite{zhang2017beyond}      & 23.66 & 0.583 & 32.43 & 0.790 & 0.67      & 175   & 0.22    \\
        CBDNet~\cite{guo2019toward}       & 30.78 & 0.801 & 38.06 & 0.942 & 4.37      & 161   & 0.19    \\
        RIDNet~\cite{Anwar_2019_ICCV}     & 38.71 & 0.951 & 39.26 & 0.953 & 1.5       & 393   & 0.68    \\
        AINDNet~\cite{kim2020transfer}    & 38.95 & 0.952 & 39.37 & 0.951 & 13.76     & 1284  & 0.49    \\
        VDN~\cite{yue2019variational}     & 39.23 & 0.955 & 39.38 & 0.952 & 7.81      & 168   & 0.20    \\
        SADNet~\cite{chang2020spatial}    & 39.46 & 0.957 & 39.59 & 0.952 & 4.23      & 76    & 0.22    \\
        DANet~\cite{yue2020dual}          & 39.47 & 0.957 & 39.58 & 0.955 & 9.15      & 59    & 0.12    \\
        CycleISP~\cite{zamir2020cycleisp} & 39.52 & 0.957 & 39.56 & 0.956 & 2.83      & 739   & 1.36    \\
        MPRNet~\cite{zamir2021multi}      & 39.62 & 0.958 & 39.80 & 0.954 & 15.74     & 2296  & 2.75    \\
        VIRNet (Ours)                     & 39.63 & 0.958 & 39.83 & 0.954 & 15.40     & 658   & 0.88    \\
        \Xhline{0.4pt}
        \textcolor[gray]{0.5}{PNGAN}~\cite{cai2021learning}
                                          & \textcolor[gray]{0.5}{40.06} & \textcolor[gray]{0.5}{0.960}
                                          & \textcolor[gray]{0.5}{40.25} & \textcolor[gray]{0.5}{0.962}
                                          & \textcolor[gray]{0.5}{15.74} & \textcolor[gray]{0.5}{2296}
                                          & \textcolor[gray]{0.5}{2.75} \\
        \Xhline{0.8pt}
    \end{tabular}
    \vspace{-3mm}
\end{table}

We compared VIRNet with several typical real-world denoising methods, including MPRNet~\cite{zamir2021multi}, CycleISP~\cite{zamir2020cycleisp}, DANet~\cite{yue2020dual}, SADNet~\cite{chang2020spatial}, VDN~\cite{yue2019variational} and so on (see Table~\ref{tab:DN_real}). To the best of our knowledge, current SotA method on these two benchmarks is PNGAN~\cite{cai2021learning}. This work, however, mainly focuses on simulating the camera pipeline to generate large amount of image pairs as training data so as to further improve the performance, instead of designing more effective denoising algorithm. Its denoiser architecture and the training strategy completely follow MPRNet. Therefore we mainly compare with MPRNet in this work.

In order to comprehensively evaluate all competing methods, Table~\ref{tab:DN_real} lists the denoising performance, as well as the model profiles, including the number of network parameters, the FLOPs, and the feedforward running time of the denoisers. The FLOPs and running time are both counted on images with a size of $512 \times 512$. From the perspective of denoising performance, the proposed VIRNet achieves a slight performance improvement compared with current SotA method MPRNet, indicating its effectiveness. However, VIRNet is of pronounced superiority in terms of model profiles, especially in the comparisons of FLOPs and running time, which more faithfully reflect the relative efficiency of our method. To intuitively compare the denoising results, we visualize two typical real examples in Fig.~\ref{fig:DN_real}, which are consistent with the quantitative results in Table~\ref{tab:DN_real}.
\begin{figure}[t]
    \centering
    \includegraphics[scale=1.14]{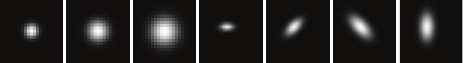}
    \vspace{-8mm}
    \caption{Seven Gaussian kernels used to generate the LR images in the synthetical super-resolution experiments.}
    \label{fig:kernels_SR}
\end{figure}
\begin{table*}[t]
    \centering
    \caption{Quantitative comparisons of different methods under a scale factor of 4. The PSNR/SSIM/LPIPS values in this table are all averaged
        over the seven kernels as shown in Fig.~\ref{fig:kernels_SR}. The results of the non-blind methods that rely on the pre-given ground truth blur
        kernel and noise level are marked in \textcolor[gray]{0.5}{gray} color to denote unfair comparisons. Besides, some
        commonly considered model profiles, namely, the number of learnable parameters (in M), the FLOPs (in G), and the running time (in seconds),
        are also listed for more comprehensive comparison. Note that the FLOPs and running time are calculated in the case of super-resolving
        the low-resolution images with a size of $256 \times 256$.} \label{tab:sisr_sf4}
    \vspace{-4mm}
    \begin{tabular}{@{}C{3.4cm}@{}|@{}C{1.0cm}@{}|@{}C{1.15cm}@{}@{}C{1.15cm}@{}@{}C{1.15cm}@{}|
                                                  @{}C{1.15cm}@{}@{}C{1.15cm}@{}@{}C{1.15cm}@{}|
                                                  @{}C{1.15cm}@{}@{}C{1.15cm}@{}@{}C{1.15cm}@{}|
                                                  @{}C{1.15cm}@{}@{}C{1.15cm}@{}@{}C{1.15cm}@{}}
        \Xhline{0.8pt}
        \multirow{2}*{Methods} & \multirow{2}*{\makecell{Noise\\Level}} & \multicolumn{3}{c|}{Set14}
                                                                        & \multicolumn{3}{c|}{CBSD68}
                                                                        & \multicolumn{3}{c|}{DIV2K100}
                                                                        & \multicolumn{3}{c}{Model Profile}\\
        \Xcline{3-14}{0.4pt}
                 &    & PSNR$\uparrow$  & SSIM$\uparrow$  & LPIPS$\downarrow$  & PSNR$\uparrow$  & SSIM$\uparrow$  & LPIPS$\downarrow$
                      & PSNR$\uparrow$  & SSIM$\uparrow$  & LPIPS$\downarrow$  & \# Param        & FLOPs           & Times \\
        \hline
        Bicubic  & \multirow{9}*{0.10}  & 24.54       & 0.6352       & 0.5257          & 24.68           & 0.6144      & 0.6044
                                     & 25.50       & 0.6762       & 0.5349          & -               & -           & 0.006     \\
        HAN~\cite{niu2020single}
                 &                   & 25.36       & 0.6731       & 0.4693          & 25.35           & 0.6494      & 0.5477
                                     & 26.30       & 0.7115       & 0.4775          & 16.07           & 4653        & 4.933     \\
        IKC~\cite{gu2019blind}
                 &                   & 27.24       & 0.7388       & 0.3534          & 26.70           & 0.7105      & 0.4253
                                     & 28.00       & 0.7741       & 0.3517          & 9.05            & 11537       & 7.392     \\
        DAN~\cite{huang2020unfolding}
                 &                   & 27.49       & 0.7464       & 0.3442          & 26.96           & 0.7204      & 0.4053
                                     & 28.31       & 0.7832       & 0.3354          & 4.33            & 5013        & 2.012     \\
        DASR~\cite{wang2021unsupervised}
                 &                   & 27.74       & 0.7512       & 0.3314          & 27.12           & 0.7231      & 0.4026
                                     & 28.32       & 0.7805       & 0.3380          & 7.25            & 839         & 0.525     \\
        BSRNet~\cite{zhang2021designing}
                 &                   & 26.84       & 0.7129       & 0.3819          & 26.38           & 0.6860      & 0.4540
                                     & 27.49       & 0.7503       & 0.3829          & 16.70           & 4706        & 1.557     \\
        VIRNet (Ours)
                 &                   & 27.89       & 0.7573       & 0.3165          & 27.27           & 0.7330      & 0.3868
                                     & 28.60       & 0.7919       & 0.3165          & 5.72            & 370         & 0.161     \\
        \Xcline{3-14}{0.4pt}
        GT+SRMD~\cite{zhang2018learning}
                 &    & \textcolor[gray]{0.5}{27.83}           & \textcolor[gray]{0.5}{0.7587}         & \textcolor[gray]{0.5}{0.3180}
                      & \textcolor[gray]{0.5}{27.12}           & \textcolor[gray]{0.5}{0.7283}         & \textcolor[gray]{0.5}{0.3962}
                      & \textcolor[gray]{0.5}{28.44}           & \textcolor[gray]{0.5}{0.7874}         & \textcolor[gray]{0.5}{0.3250}
                      & \textcolor[gray]{0.5}{1.55 }           & \textcolor[gray]{0.5}{407   }         & \textcolor[gray]{0.5}{0.106}     \\
        GT+USRNet~\cite{zhang2020deep}
                 &    & \textcolor[gray]{0.5}{27.90}           & \textcolor[gray]{0.5}{0.7747}         & \textcolor[gray]{0.5}{0.3181}
                      & \textcolor[gray]{0.5}{26.88}           & \textcolor[gray]{0.5}{0.7408}         & \textcolor[gray]{0.5}{0.3988}
                      & \textcolor[gray]{0.5}{28.84}           & \textcolor[gray]{0.5}{0.8086}         & \textcolor[gray]{0.5}{0.3042}
                      & \textcolor[gray]{0.5}{17.20}           & \textcolor[gray]{0.5}{38893 }         & \textcolor[gray]{0.5}{9.214 }   \\
        \hline \hline
        Bicubic  & \multirow{9}*{2.55}  & 24.51           & 0.6314     & 0.5590     & 24.67     & 0.6114       & 0.6332
                                        & 25.44           & 0.6713     & 0.5777     &  -        & -            & 0.006     \\
        DnCNN~\cite{zhang2017beyond}+HAN~\cite{niu2020single}
                 &                      & 25.24           & 0.6596     & 0.5213     & 25.23     & 0.6350       & 0.6024
                                        & 26.11           & 0.6925     & 0.5618     & 16.07     & 4653         & 4.933     \\
        DnCNN~\cite{zhang2017beyond}+IKC~\cite{gu2019blind}
                 &                      & 25.80           & 0.6795     & 0.4590     & 25.66     & 0.6546       & 0.5297
                                        & 26.59           & 0.7144     & 0.4702     & 9.05      & 11537        & 7.392     \\
        DnCNN~\cite{zhang2017beyond}+DAN~\cite{huang2020unfolding}
                 &                      & 25.43           & 0.6685     & 0.4873     & 25.46     & 0.6475       & 0.5524
                                        & 26.32           & 0.7062     & 0.4964     & 4.33      & 5013         & 2.012     \\
        DASR~\cite{wang2021unsupervised}
                 &                      & 27.05           & 0.7197     & 0.3672     & 26.50     & 0.6881       & 0.4479
                                        & 27.58           & 0.7491     & 0.3830     & 7.25      & 839          & 0.525     \\
        BSRNet~\cite{zhang2021designing}
                 &                      & 26.70           & 0.7093     & 0.3801     & 26.23     & 0.6797       & 0.4562
                                        & 27.31           & 0.7431     & 0.3858     & 16.70     & 4706         & 1.557     \\
        VIRNet (Ours)
                 &                      & 27.18           & 0.7255     & 0.3524     & 26.63     & 0.6975       & 0.4305
                                        & 27.86           & 0.7597     & 0.3588     & 5.72      & 370          & 0.161     \\
        \Xcline{3-14}{0.4pt}
        GT+SRMD~\cite{zhang2018learning}
                 &       & \textcolor[gray]{0.5}{27.15}           & \textcolor[gray]{0.5}{0.7270}          & \textcolor[gray]{0.5}{0.3549}
                         & \textcolor[gray]{0.5}{26.54}           & \textcolor[gray]{0.5}{0.6954}          & \textcolor[gray]{0.5}{0.4367}
                         & \textcolor[gray]{0.5}{27.76}           & \textcolor[gray]{0.5}{0.7578}          & \textcolor[gray]{0.5}{0.3647}
                         & \textcolor[gray]{0.5}{1.55 }           & \textcolor[gray]{0.5}{407   }          & \textcolor[gray]{0.5}{0.106 }    \\
        GT+USRNet~\cite{zhang2020deep}
                 &       & \textcolor[gray]{0.5}{27.42}           & \textcolor[gray]{0.5}{0.7484}          & \textcolor[gray]{0.5}{0.3395}
                         & \textcolor[gray]{0.5}{26.53}           & \textcolor[gray]{0.5}{0.7125}          & \textcolor[gray]{0.5}{0.4249}
                         & \textcolor[gray]{0.5}{28.27}           & \textcolor[gray]{0.5}{0.7809}          & \textcolor[gray]{0.5}{0.3366}
                         & \textcolor[gray]{0.5}{17.20}           & \textcolor[gray]{0.5}{38893 }          & \textcolor[gray]{0.5}{9.214 }    \\
        \hline \hline
        Bicubic  & \multirow{9}*{7.65} & 24.25           & 0.6051          & 0.6749             & 24.38           & 0.5836         & 0.7662
                                       & 25.08           & 0.6381          & 0.7460             &  -              & -              & 0.006     \\
        DnCNN~\cite{zhang2017beyond}+HAN~\cite{niu2020single}
                 &                     & 24.30           & 0.5623          & 0.6899             & 24.24           & 0.5321         & 0.7934
                                       & 24.78           & 0.5613          & 0.7837             & 16.07           & 4653           & 4.933     \\
        DnCNN~\cite{zhang2017beyond}+IKC~\cite{gu2019blind}
                 &                     & 25.37           & 0.6566          & 0.4639             & 25.28           & 0.6323         & 0.5351
                                       & 26.13           & 0.6935          & 0.4738             & 9.05            & 11537          & 7.392     \\
        DnCNN~\cite{zhang2017beyond}+DAN~\cite{huang2020unfolding}
                 &                     & 25.05           & 0.6474          & 0.4991             & 25.12           & 0.6253         & 0.5702
                                       & 25.97           & 0.6869          & 0.5094             & 4.33            & 5013           & 2.012     \\
        DASR~\cite{wang2021unsupervised}
                 &                     & 26.19           & 0.6845          & 0.4055             & 25.75           & 0.6516         & 0.4935
                                       & 26.75           & 0.7165          & 0.4229             & 7.25            & 839            & 0.525     \\
        BSRNet~\cite{zhang2021designing}
                 &                     & 25.58           & 0.6703          & 0.4195             & 25.20           & 0.6385         & 0.5083
                                       & 26.08           & 0.7023          & 0.4442             & 16.70           & 4706           & 1.557     \\
        VIRNet (Ours)
                 &                     & 26.22           & 0.6873          & 0.3973             & 25.81           & 0.6576         & 0.4830
                                       & 26.91           & 0.7227          & 0.4103             & 5.72            & 370            & 0.161     \\
        \Xcline{3-14}{0.4pt}
        GT+SRMD~\cite{zhang2018learning}
                 &                     & \textcolor[gray]{0.5}{26.20}      & \textcolor[gray]{0.5}{0.6881}        & \textcolor[gray]{0.5}{0.3999}
                                       & \textcolor[gray]{0.5}{25.74}      & \textcolor[gray]{0.5}{0.6556}        & \textcolor[gray]{0.5}{0.4893}
                                       & \textcolor[gray]{0.5}{26.83}      & \textcolor[gray]{0.5}{0.7212}        & \textcolor[gray]{0.5}{0.4156}
                                       & \textcolor[gray]{0.5}{1.55 }      & \textcolor[gray]{0.5}{407   }        & \textcolor[gray]{0.5}{0.106 }    \\
        GT+USRNet~\cite{zhang2020deep}
                 &                     & \textcolor[gray]{0.5}{26.81}      & \textcolor[gray]{0.5}{0.7115}        & \textcolor[gray]{0.5}{0.3705}
                                       & \textcolor[gray]{0.5}{26.09}      & \textcolor[gray]{0.5}{0.6745}        & \textcolor[gray]{0.5}{0.4616}
                                       & \textcolor[gray]{0.5}{27.43}      & \textcolor[gray]{0.5}{0.7440}        & \textcolor[gray]{0.5}{0.3818}
                                       & \textcolor[gray]{0.5}{17.20}      & \textcolor[gray]{0.5}{38893 }        & \textcolor[gray]{0.5}{9.214 }    \\
        \Xhline{0.8pt}
    \end{tabular}
    \vspace{-2mm}
\end{table*}
\begin{figure*}[t]
    \centering
    \includegraphics[scale=0.92]{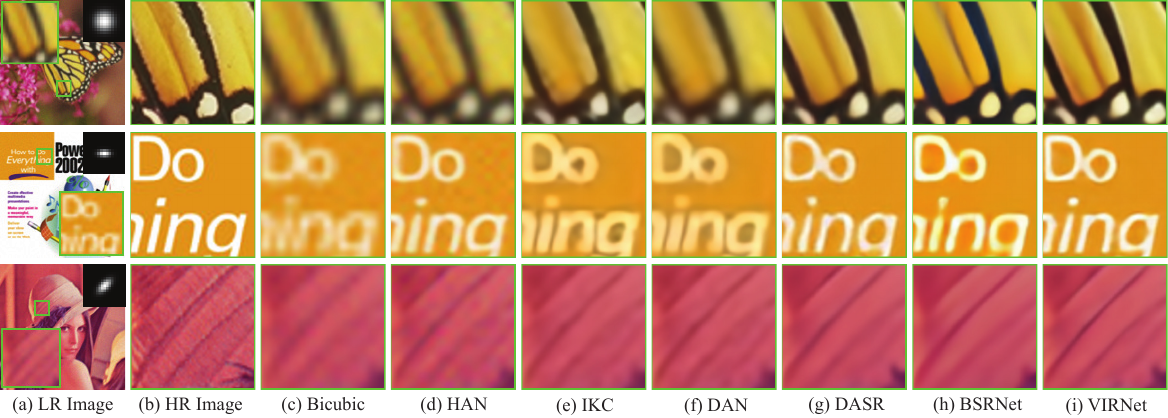}
    \vspace{-7mm}
    \caption{Visualized super-resolution results of different methods on a synthetic example of Set14~\cite{Zeyde2010} under a scale factor of 4. Specifically,
        the noise level is 2.55, and the blur kernel is shown on the upper-right corner  of the LR image.}
    \label{fig:sisr_syn}
\end{figure*}

\subsection{Image Super-resolution Experiments} \label{subsec:sr-experiment}
In this section, we apply our proposed VIRNet to blind image super-resolution. Following~\cite{wang2021unsupervised}, the DF2K dataset (containing 800 images from DIV2K~\cite{Agustsson_2017_CVPR_Workshops} and 2650 images from Flickr2K~\cite{Timofte_2017_CVPR_Workshops}) was employed as our training data. When synthesizing the LR images, we followed the settings of current blind SR literature~\cite{huang2020unfolding,wang2021unsupervised}, i.e.,
\begin{equation}
    \bm{y} = (\bm{z}\otimes \bm{k})\downarrow_{s}^d + \bm{n},
    \label{eq:degrade-sisr}
\end{equation}
where $\bm{y}$ and $\bm{z}$ denote the low-resolution and high-resolution image respectively, $\otimes$ is the 2-D convolution, $\downarrow_{s}^d$ is the direct\footnote{Extracting the upper-left pixel for each $p\times p$ patch.} downsampler with a scale factor of $s$, and $\bm{n}$ is the i.i.d. Gaussian noise with noise level $\sigma$. For the blur kernel $\bm{k}$, we adopted the general anisotropic Gaussian kernel with a size of $21 \times 21$, and its covariance matrix $\bm{\Sigma}$ was generated like~\cite{Shocher2018} during training, i.e.,
\begin{equation}
    \bm{U} = \begin{bmatrix}
        \cos{\theta} & -\sin{\theta} \\
        \sin{\theta} & \cos{\theta}
    \end{bmatrix}, ~
    \bm{\Lambda} = \begin{bmatrix}
        l_1^2  & 0 \\
        0    & l_2^2
    \end{bmatrix}, ~
    \bm{\Sigma} = \bm{U}\bm{\Lambda}\bm{U}^T.
    \label{eq:sigma-anisotropic}
\end{equation}
To be specific, $l_1$, $l_2$, and $\theta$ are randomly sampled from $[0, s]$, $[0, s]$, and $[0, \pi]$, respectively. According to $\bm{\Lambda}$, it is easy to calculate the parameters of $\lambda_1$, $\lambda_2$, and $\rho$ in Eq.~\eqref{eq:kennel_define}.
For the noise level $\sigma$, we set its range to be $[0,15]$.
\begin{figure*}[t]
    \centering
    \includegraphics[scale=1.04]{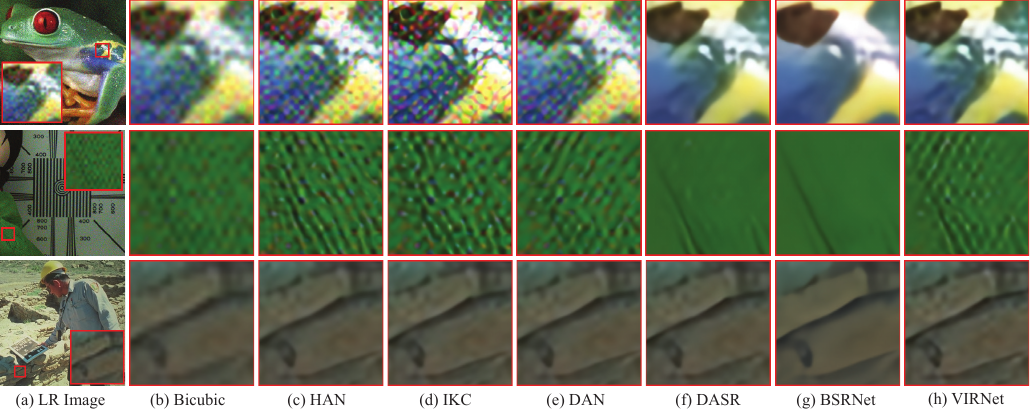}
    \vspace{-7mm}
    \caption{Visualized super-resolution results of different methods on three examples of RealSRSet~\cite{zhang2021designing} under
    a scale factor of 4.}
    \label{fig:sisr_real}
    \vspace{-3mm}
\end{figure*}
\begin{figure*}[t]
    \centering
    \includegraphics[scale=1.05]{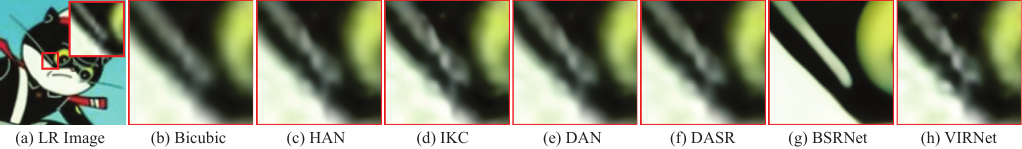}
    \vspace{-8mm}
    \caption{One typical super-resolution result on an example from RealSRSet~\cite{zhang2021designing} with a scale factor of 4. In this example, most of
    the methods fail to obtain one high-quality HR image except for BSRNet, mainly owning to their mismatch on image degradation.}
    \label{fig:sisr_real_fail}
    \vspace{-2mm}
\end{figure*}

\subsubsection{Results on Synthetic Data}
To be capable of quantitatively evaluating different methods, we first conducted some synthetic experiments on three commonly used datasets, including Set14~\cite{Zeyde2010}, CBSD68~\cite{arbelaez2010contour}, and DIV2K100 (the validation set of DIV2K~\cite{Agustsson_2017_CVPR_Workshops}). To make a thorough comparison on various degradation, we considered seven representative and diverse kernels as shown in Fig.~\ref{fig:kernels_SR}, including three isotropic Gaussian kernels with different kernel widths (i.e., 0.4$s$, 0.6$s$, and 0.8$s$) and four anisotropic Gaussian kernels, where $s$ is the scale factor. In addition, three noise levels (i.e., $0$, $2.55$, and $7.65$) are considered following~\cite{zhang2020deep}. As for the metrics, aside from the commonly used PSNR and SSIM~\cite{Wang2004}, we also adopted LPIPS~\cite{zhang2018unreasonable} to measure the perceptual similarity. Note that PSNR and SSIM are calculated on the Y channel of the YCbCr space, while LPIPS is calculated in the sRGB space.
\begin{table}[t]
    \centering
    \caption{The non-reference comparison results of different methods on the real-world dataset SRRealSet~\cite{zhang2021designing} under a scale
        factor of 4. The best and second best results are highlighted by \textbf{bold} and \underline{underline}, respectively.} \label{tab:non_reference_real}
    \vspace{-4mm}
    \begin{tabular}{@{}C{1.3cm}@{}|@{}C{1.2cm}@{}@{}C{1.0cm}@{}@{}C{0.9cm}@{}@{}C{0.9cm}@{}
                                   @{}C{1.2cm}@{}@{}C{1.2cm}@{}@{}C{1.2cm}@{}@{}C{1.4cm}@{}}
        \Xhline{0.8pt}
        \multirow{2}*{Metrics} & \multicolumn{7}{c}{Methods} \\
        \Xcline{2-8}{0.4pt}
                               & Bicubic  & HAN            & IKC            & DAN  & DASR  & BSRNet             & VIRNet            \\
        \Xhline{0.4pt}
        NRQM$\uparrow$         & 3.35     & 4.53           & \textbf{4.72}  & 4.43  & 4.18  & 4.30              & \underline{4.68}  \\
        \Xhline{0.4pt}
        PI$\downarrow$         & 6.36     & 5.55           & \textbf{5.33}  & 5.54  & 5.73  & 5.50              & \underline{5.39}     \\
        \Xhline{0.8pt}
    \end{tabular}
    \vspace{-3mm}
\end{table}

We consider three categories of comparison methods: 1) classical Bicubic interpolation method; 2) five blind super-resolution methods, including HAN~\cite{niu2020single}, IKC~\cite{gu2019blind}, DAN~\cite{huang2020unfolding}, DASR~\cite{wang2021unsupervised}, and BSRNet~\cite{zhang2021designing}; 3) two non-blind methods, i.e., SRMD~\cite{zhang2018learning} and USRNet~\cite{zhang2020deep}, which rely on the pre-given blur kernel and noise level as input. For these non-blind methods, we provided the ground truth blur kernel and noise level for them and denoted their results with the format of ``GT+X'' (e.g., GT+SRMD). In addition, for the methods of HAN, IKC and DAN, we firstly denoised the noisy low-resolution image using DnCNN~\cite{zhang2017beyond} and then super-resolved it in the cases of $\sigma=\{2.55, 7.65\}$, since these methods do not consider image noise during training.

Table~\ref{tab:sisr_sf4} lists the comparison results of different methods under a scale factor of 4, and more results under scale factors of 2 and 3 are put
into Appendix~\ref{subsec:exp-sr-supp}. From Table~\ref{tab:sisr_sf4}, it can be seen that the proposed VIRNet achieves the best performance among
blind methods in all cases. Especially, compared with the non-blind methods, VIRNet is still able to obtain slightly better or at least comparable results even though they make use of the groud truth information of blur kernel and noise level. This indicates the effectiveness of the proposed blind framework which is capable of handling the tasks of degradation estimation and image restoration simultaneously. Further, taking the model profiles into consideration, the superiority of VIRNet is more evident. Specifically, VIRNet has fewer number of parameters, fewer FLOPs, and faster speed than both the SotA blind method DASR~\cite{wang2021unsupervised} and the non-blind method USRNet~\cite{zhang2020deep}.

In Fig.~\ref{fig:sisr_syn}, we display the denoising results on three typical visual examples of Set14 with a scale factor of 4. Note that we only show the results of blind super-resolution methods for a fair comparison. It can be easily seen that the proposed VIRNet is able to recover more realistic and sharper results, which are evidently closer to the groud truth high-resolution images than other methods. The results of most comparison methods are relatively blurry and lose some image details. In the second example (the middle row), IKC and DAN lead to a relatively severe corruption of the original image color. That's possibly caused by the inconsistency of their multiple iterations, since they both adopt the coarse-to-fine manner to gradually adjust the results. Due to the careful considerations of the degradation model, DASR and BSRNet also perform well compared with other methods. However, VIRNet still evidently surpasses them in terms of the quantitative and qualitative results. This further substantiates the effectiveness of the proposed variational framework.

\begin{table}[t]
    \centering
    \caption{The PSNR results of FFDNet~\cite{zhang2018ffdnet} with the groudtruth noise level and with that estimated by the proposed VIRNet on
    CBSD68~\cite{arbelaez2010contour} under the i.i.d. Gaussian noise setting.}
    \label{tab:iid_noise_est}
    \vspace{-4mm}
    \begin{tabular}{@{}C{2.8cm}@{}|@{}C{2.0cm}@{}@{}C{2.0cm}@{}@{}C{2.0cm}@{}}
        \Xhline{0.8pt}
        \multirow{2}*{Methods}         & \multicolumn{3}{c}{Noise Levels} \\
        \Xcline{2-4}{0.4pt}
                                       & $\sigma =15$  & $\sigma =25$  & $\sigma =50$     \\
        \Xhline{0.4pt}
        $\text{FFDNet}_{\text{GT}}$    & 33.87         & 31.20         & 27.95            \\
        \Xhline{0.4pt}
        $\text{FFDNet}_{\text{VIR}}$   & 33.87         & 31.20         & 27.92            \\
        \Xhline{0.8pt}
    \end{tabular}
    \vspace{-4mm}
\end{table}
\begin{table}[t]
    \centering
    \caption{The PSNR results of FFDNet~\cite{zhang2018ffdnet} with the groudtruth noise variance map and with that estimated by the proposed VIRNet
    on CBSD68~\cite{arbelaez2010contour} under the non-i.i.d. Gaussian noise.}
    \label{tab:noniid_noise_est}
    \vspace{-4mm}
    \begin{tabular}{@{}C{2.8cm}@{}|@{}C{2.0cm}@{}@{}C{2.0cm}@{}@{}C{2.0cm}@{}}
        \Xhline{0.8pt}
        \multirow{2}*{Methods}         & \multicolumn{3}{c}{Noise Cases} \\
        \Xcline{2-4}{0.4pt}
                                       & Case 1        & Case 2        & Case 3     \\
        \Xhline{0.4pt}
        $\text{FFDNet}_{\text{GT}}$    & 28.79         & 28.42         & 28.68            \\
        \Xhline{0.4pt}
        $\text{FFDNet}_{\text{VIR}}$   & 28.75         & 28.39         & 28.65            \\
        \Xhline{0.8pt}
    \end{tabular}
    \vspace{-3mm}
\end{table}
\begin{table}[t]
    \centering
    \caption{The average kernel MSE (1E-5) values of different methods on DIV2K20 with a scale factor of 4.}
    \label{tab:kernel_mse}
    \vspace{-4mm}
    \begin{tabular}{@{}C{1.4cm}@{}|@{}C{2.2cm}@{}@{}C{2.0cm}@{}@{}C{1.8cm}@{}@{}C{1.5cm}@{}}
        \Xhline{0.8pt}
        \multirow{2}*{Metrics} & \multicolumn{4}{c}{Methods} \\
        \Xcline{2-5}{0.4pt}
                               & KernelGan~\cite{bell2019blind}  & DIPFKP~\cite{liang2021flow}  & BSRDM~\cite{yue2020bsrdm}   & VIRNet     \\
        \Xhline{0.4pt}
        MSE$\downarrow$        & 4.34                            & 1.73                         & 1.60                        & 0.89         \\
        \Xhline{0.8pt}
    \end{tabular}
    \vspace{-3mm}
\end{table}

\subsubsection{Results on Real Data}
In this part, we further justify the effectiveness of the proposed VIRNet on the real-world dataset RealSRSet~\cite{zhang2021designing}. It contains 20 real images that are commonly used in previous literature~\cite{ignatov2017dslr,martin2001database,matsui2017sketch,zhang2018ffdnet} or downloaded from the internet. Since the underlying high-resolution images for them are not available, we thus mainly evaluate different methods by visual comparisons. Fig.~\ref{fig:sisr_real} displays three typical super-resolution examples with a scale factor of 4. In the first (top row) and second (middle row) examples, the LR images both contain some image noises, which makes the super-resolution goal more challenging. The methods of Bicubic, HAN, IKC, and DAN all fail to deal with such cases and produce some artifacts in the areas with image noises. As for DASR and BSRNet, they, unfortunately, erase the high-frequency image details when removing the image noises. One can easily observe that the proposed VIRNet makes a good trade-off between preserving the image details and removing the image noises. In the third (bottom row) example, the results of IKC and VIRNet are more natural and realistic than others that are all blurry to different extents. These results verify the stable and consistently well performance of VIRNet in the real-world super-resolution task.
\begin{table}[!t]
    \centering
    \caption{The average comparison results of USRNet combined with different kernel estimation methods on DIV2K20 with a scale factor of 4. Note that ``GT+USRNet'' denotes the results of USRNet with the pre-given groudtruth kernels.}
    \label{tab:kernel_psnr_lpips}
    \vspace{-4mm}
    \begin{tabular}{@{}C{4.4cm}@{}|@{}C{1.5cm}@{}@{}C{1.5cm}@{}@{}C{1.5cm}@{}}
        \Xhline{0.8pt}
        \multirow{2}*{Methods} & \multicolumn{3}{c}{Metrics} \\
        \Xcline{2-4}{0.4pt}
                               & PSNR$\uparrow$ & SSIM$\uparrow$  & LPIPS$\downarrow$   \\
        \Xhline{0.4pt}
        KernelGan~\cite{bell2019blind}+USRNet~\cite{zhang2020deep}
                               & 14.52          & 0.358           & 0.522        \\
        \Xhline{0.4pt}
        DIPFKP~\cite{liang2021flow}+USRNet~\cite{zhang2020deep}
                               & 24.02          & 0.676           & 0.363        \\
        \Xhline{0.4pt}
        BSRDM~\cite{yue2020bsrdm}+USRNet~\cite{zhang2020deep}
                               & 27.07          & 0.771           & 0.326        \\
        \Xhline{0.4pt}
        VIRNet+USRNet~\cite{zhang2020deep}
                               & 28.53          & 0.796           & 0.313        \\
        \Xhline{0.4pt}
        GT+USRNet~\cite{zhang2020deep}
                               & 29.19          & 0.805           & 0.302        \\
        \Xhline{0.8pt}
    \end{tabular}
    \vspace{-3mm}
\end{table}

In Table~\ref{tab:non_reference_real}, we adopt two non-reference metrics (i.e., NRQM~\cite{ma2017learning} and PI~\cite{blau20182018}) to further quantitatively evaluate different methods. It can be seen that the proposed VIRNet achieves the second best results in terms of both metrics, only slightly worse than IKC, which indicates the effectiveness of our method. Combining its better visual performance as shown in Fig.~\ref{fig:sisr_real}, especially compared with those obtained by IKC, the relative superiority of the proposed method can still be validated.

\subsubsection{Discussion on Blur Kernel}
How to properly and generally set the degradation model, especially the blur kernel, is still an open yet challenging research topic in the field of image super-resolution~\cite{zhang2021designing,wang2021real}. In this work, we adopt the anisotropic Gaussian assumption for the blur kernel following most of current related literature ~\cite{Riegler2015,zhang2018learning,gu2019blind,huang2020unfolding,wang2021unsupervised,liang2021flow}. Even though such general kernel hypothesis is sufficient in most of scenarios, it still possibly leads to some unsatisfactory results in some cases. For instance, Fig.~\ref{fig:sisr_real_fail} shows one typical failed example. In this example, the LR image contains some obvious ``ringing artifacts'' that look like bands or ghosts near edges, which are usually produced by the sharping algorithm or image compression. The Gaussian kernel-based methods (i.e., IKC~\cite{gu2019blind}, DAN~\cite{huang2020unfolding}, DASR~\cite{wang2021unsupervised}, and the proposed VIRNet) all cannot finely resolve this image, while BSRNet~\cite{zhang2021designing} performs well in this case, mainly because it integrates multiple complicated kernel settings. Therefore, it is necessary to exploit a more rational and general kernel (or degradation) modeling method for image super-resolution, and we leave this to future work.
\begin{table*}[t]
    \centering
    \caption{The qualitative comparisons of different posteriori factorizations on DIV2K100 with scale factor 4. Note that these results
            are averaged on the seven kernels shown in Fig.~\ref{fig:kernels_SR}, and the noise level is set as $2.55$.}
    \label{tab:posterior_analysis}
    \vspace{-4mm}
    \begin{tabular}{@{}C{2.4cm}@{}|@{}C{7.2cm}@{}|@{}C{2.0cm}@{}@{}C{1.8cm}@{}@{}C{1.5cm}@{}}
        \Xhline{0.8pt}
        \multirow{2}*{Models}  &\multirow{2}*{Posteriori Factorizations}& \multicolumn{3}{c}{Metrics} \\
        \Xcline{3-5}{0.4pt}
                                  &        & PSNR$\uparrow$  & SSIM$\uparrow$   & LPIPS$\downarrow$     \\
        \Xhline{0.4pt}
        \textit{Baseline1}
            & $q(\bm{z},\bm{\sigma}^2,\bm{\varLambda}|\bm{y}) = q(\bm{z}|\bm{y})q(\bm{\sigma}^2|\bm{y})q(\bm{\varLambda}|\bm{y})$
                                           & 27.69           &0.7568            &0.3641 \\
        \Xhline{0.4pt}
        \textit{Baseline2}
            & $q(\bm{z},\bm{\sigma}^2,\bm{\varLambda}|\bm{y}) = q(\bm{z}|\bm{y},\bm{\sigma}^2)q(\bm{\sigma}^2|\bm{y})q(\bm{\varLambda}|\bm{y})$
                                           & 27.71           &0.7561            &0.3632 \\
        \Xhline{0.4pt}
        \textit{Baseline3}
            & $q(\bm{z},\bm{\sigma}^2,\bm{\varLambda}|\bm{y}) = q(\bm{z}|\bm{y},\bm{\varLambda})q(\bm{\sigma}^2|\bm{y})q(\bm{\varLambda}|\bm{y})$
                                           & 27.86           &0.7596            &0.3611 \\
        \Xhline{0.4pt}
        VIRNet
            & $q(\bm{z},\bm{\sigma}^2,\bm{\varLambda}|\bm{y}) = q(\bm{z}|\bm{y},\bm{\sigma}^2,\bm{\varLambda})q(\bm{\sigma}^2|\bm{y})q(\bm{\varLambda}|\bm{y})$
                                           & 27.86           & 0.7597           &0.3588 \\
        \Xhline{0.8pt}
    \end{tabular}
    \vspace{-3mm}
\end{table*}

\subsection{Degradation Estimation Experiments} \label{subsec:degradation-est-experiment}
In this subsection, we empirically verify the effectiveness of our method in the task of degradation estimation, including noise estimation and kernel estimation.

\subsubsection{Noise Estimation} \label{subsubsec:noise_est}
Different from most of the current IR methods, the pixel-wise non-i.i.d. Gaussian assumption is adopted to fit the noise distribution
in our method. Next, we analyze the performance of our method with such an assumption under several common noise types in IR tasks:

\vspace{1mm}
\noindent\textbf{I.I.D. Gaussian Noise.} Even though VIRNet is designed based on non-i.i.d. Gaussian noise assumption, it can be generalized well to the i.i.d. Gaussian noise as shown in Table~\ref{tab:psnr_iidawgn}. To further quantitatively illustrate this point, we take the estimated noise level by our method as the input of FFDNet~\cite{zhang2018ffdnet}, which is a typical non-blind i.i.d. Gaussian denoising method that relies on the pre-known noise level. Table~\ref{tab:iid_noise_est} lists the PSNR comparison results of FFDNet with different noise level settings, in which $\text{FFDNet}_{\text{VIR}}$ and $\text{FFDNet}_{\text{GT}}$ denote the results of FFDNet taking the predicted noise level by VIRNet and the ground truth noise level as input respectively. We can see that $\text{FFDNet}_{\text{VIR}}$ is able to achieve the same performance with $\text{FFDNet}_{\text{GT}}$ when $\sigma = \{15, 25\}$, or very close performance when $\sigma = 50$, even though $\text{FFDNet}_{\text{GT}}$ makes use of the true noise level. This indicates that VIRNet is capable of properly estimating the noise levels of the i.i.d. Gaussian noise.
\begin{figure}[!t]
    \centering
    \includegraphics[scale=0.515]{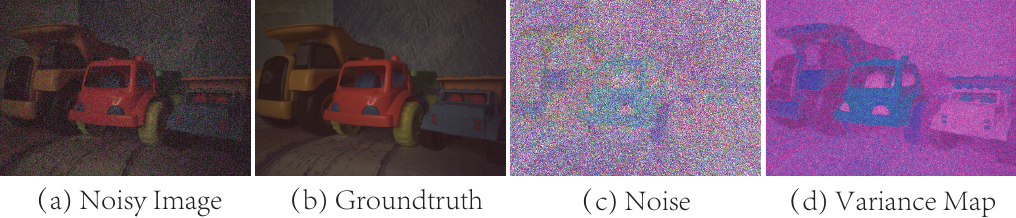}
    \vspace{-7.0mm}
    \caption{One typical visual example in SIDD~\cite{Abdelhamed2018} dataset. From left to right: (a) noisy image $\bm{y}$; (b) noise-free image $\bm{x}$; (c) image noise calculated through $|\bm{y}-\bm{x}|$; (d) variance map predicted by the proposed method.}
    \label{fig:noise-var}
    \vspace{-2mm}
\end{figure}

\vspace{1mm}
\noindent\textbf{Non-I.I.D. Gaussian Noise.} In Sec.~\ref{subsubsec:synthetica-denoising}, we adopt three groups of noise variance maps (see Fig.~\ref{fig:sigmaMap} (b1-d1)) to synthesize the testing data, to evaluate the performance of VIRNet under the non-i.i.d. Gaussian noise. Correspondingly, Fig.~\ref{fig:sigmaMap} (b2-d2) further displays the variance maps predicted by VIRNet for easy visualization. It can be seen that these predicted variance maps have very similar spatial variation with the ground truth ones, which are expected to facilitate the subsequent denoising task or other non-blind denoising methods. To justify this point, we also apply these predicted variance maps in FFDNet~\cite{zhang2018ffdnet} to test its performance under the non-i.i.d. Gaussian noise, and the quantitative comparisons are listed in Table~\ref{tab:noniid_noise_est}. One can see that $\text{FFDNet}_{\text{VIR}}$ and $\text{FFDNet}_{\text{GT}}$ have very similar performance, and the performance difference between them is less than 0.04dB PSNR. This indicates that VIRNet is able to effectively handle such complicated noise distribution.

\vspace{1mm}
\noindent\textbf{Signal-dependent Noise.} The challenge in real-world image denoising is mainly attributed to the signal-dependentness of the image noise. Fig.~\ref{fig:noise-var} shows one typical real-world noisy example coming from SIDD~\cite{Abdelhamed2018} dataset and the corresponding variance maps predicted by VIRNet. Note that the variance map has been enlarged several times for easy visualization. It is easy to see that the estimated noise variance map depicts strong relevance to the pixel illumination, implying that the proposed VIRNet is able to finely approximate the signal-dependent real noise.

\subsubsection{Kernel Estimation}
As is well known, kernel estimation plays an important role in blind image super-resolution~\cite{efrat2013accurate}. To evaluate the effectiveness of VIRNet in this subtask, we compare three recent kernel estimation methods specifically designed for super-resolution, including KernelGan~\cite{bell2019blind}, DIPFKP~\cite{liang2021flow}, and BSRDM~\cite{yue2020bsrdm}. Since these three methods are all relatively time-consuming, we randomly select 20 images from the validation set of DIV2K~\cite{Agustsson_2017_CVPR_Workshops} (denoted as DIV2K20) as testing data. The LR images are synthesized using the last four anisotropic Gaussian kernels in Fig.~\ref{fig:kernels_SR} under a scale factor of 4, and the noise level is set as $2.55$.

As for the evaluation, we use two ways to compare the performance of different methods. Firstly, the MSE between the estimated kernel and the ground truth kernel is an intuitive metric that directly reflects the accuracy of the estimated kernel. The detailed comparison results are listed in Table~\ref{tab:kernel_mse}. Secondly, we apply the estimated blur kernels in a non-blind super-resolution method USRNet~\cite{zhang2020deep} and then compare the recovered HR image in terms of PSNR, SSIM~\cite{Wang2004}, and LPIPS~\cite{zhang2018unreasonable}. The comparison results on these three metrics are listed in Table~\ref{tab:kernel_psnr_lpips}. From both tables, one can easily observe that the proposed VIRNet exhibits evident superiority over other competing methods.

\subsection{Ablation Study} \label{subsec:ablation-exp}
\subsubsection{Posteriori Factorization} \label{subsubsec:model-analysis-experiment}
When designing the variational inference algorithm in Sec.~\ref{sec:inference}, we factorize the variational distribution $q(\bm{z},\bm{\sigma}^2,\bm{\varLambda}|\bm{y})$ into a conditional format of Eq.~\eqref{eq:poster-factorization}, which fundamentally induces the cascaded inference framework in Fig.~\ref{fig:framework}. In fact, different factorized assumptions on $q(\bm{z},\bm{\sigma}^2,\bm{\varLambda}|\bm{y})$ will lead to different designs on the inference framework. For example, the following unconditional factorization
\begin{equation}
    q(\bm{z},\bm{\sigma}^2,\bm{\varLambda}|\bm{y}) = q(\bm{z}|\bm{y}) q(\bm{\sigma}^2|\bm{y}) q(\bm{\varLambda}|\bm{y}),
    \label{eq:poster-factorization-uncoditional}
\end{equation}
will induce a parallel inference architecture. Specifically, the three sub-networks, namely \textit{SNet}, \textit{KNet}, and \textit{RNet}, will
feedforward independently in such a parallel framework, but they can interact during backpropagation through ELBO. Please refer to our previous
conference version~\cite{yue2019variational} for a thorough overview on this point.

To validate the superiority of the conditional form of Eq.~\eqref{eq:poster-factorization}, we consider different posteriori factorization and empirically compare their performance on the task of image super-resolution, since it involves more general degradation than image denoising. Table~\ref{tab:posterior_analysis} lists the average comparison results on DIV2K100 with a scale factor of 4. As compared with \textit{Baseline1}, it can be seen that VIRNet achieves evident performance gain, which indicates that the degradation information (i.e., the noise level and the blur kernel) can facilitate the image restoration task. In fact, such a conditional factorization in VIRNet is consistent with the classical model-based methods that decompose the blind IR in two subproblems, namely degradation estimation and image restoration. The superiority of \textit{Baseline3} over \textit{Baseline2} demonstrates that the kernel information can bring up more marginal performance improvement than the noise level, complying with the conclusion in~\cite{efrat2013accurate}. However, the performance gain of VIRNet over \textit{Baseline3} on LPIPS substantiates that conditioning on the noise level can further improve the perceptual quality of the recovered images.
\begin{table}[t]
    \centering
    \caption{Performance of VIRNet under different $\varepsilon_0$ values on SIDD validation set.} \label{tab:hyper_epsilon}
    \vspace{-4mm}
    \begin{tabular}{@{}C{1.2cm}@{}|@{}C{1.10cm}@{}@{}C{1.10cm}@{}@{}C{1.10cm}@{}@{}C{1.10cm}@{}@{}C{1.10cm}@{}@{}C{1.10cm}@{}@{}C{1.0cm}@{}}
        \Xhline{0.8pt}
          \multirow{2}*{Metrics}     & \multicolumn{7}{c}{$\varepsilon_0^2$}   \\
        \Xcline{2-8}{0.4pt}
                           & 1e-4    &1e-5      &1e-6     &1e-7    &1e-8     & 1e-9      &MSE      \\
        \Xhline{0.4pt}
          PSNR             & 39.34   &39.57     & 39.63   &39.59   & 39.57   &39.53      & 39.51   \\
          SSIM             & 0.9167  & 0.9192   & 0.9194  &0.9193  & 0.9191  &0.9188     &0.9187  \\
        \Xhline{0.8pt}
    \end{tabular}
    \vspace{-3mm}
\end{table}
\begin{table}[t]
    \centering
    \caption{Performance of VIRNet under different $p$ values on SIDD validation set.} \label{tab:hyper_p}
    \vspace{-4mm}
    \begin{tabular}{@{}C{1.8cm}@{}|@{}C{1.15cm}@{}@{}C{1.15cm}@{}@{}C{1.15cm}@{}@{}C{1.15cm}@{}@{}C{1.15cm}@{}@{}C{1.15cm}@{}}
        \Xhline{0.8pt}
          \multirow{2}*{Metrics}     & \multicolumn{6}{c}{$p$}   \\
        \Xcline{2-7}{0.4pt}
                           &3        & 5       &7        &9        &11       &15       \\
        \Xhline{0.4pt}
          PSNR             & 39.61   & 39.61   &39.63    &39.61    &39.62    & 39.62       \\
          SSIM             & 0.9195  & 0.9196  &0.9194   &0.9195   &0.9195   & 0.9195    \\
        \Xhline{0.8pt}
    \end{tabular}
    \vspace{-2mm}
\end{table}
\begin{table}[!t]
    \centering
    \caption{Quantitative comparison of VIRNet and MSE loss with different restoration backbones on SIDD validation set.} \label{tab:backbone_ablation}
    \vspace{-4mm}
    \begin{tabular}{@{}C{1.4cm}@{}|@{}C{1.4cm}@{}|@{}C{1.50cm}@{}@{}C{1.40cm}@{}@{}C{1.50cm}@{}@{}C{1.50cm}@{}}
        \Xhline{0.8pt}
          \multirow{2}*{Methods} & \multirow{2}*{Metrics}     & \multicolumn{4}{c}{Backbones}   \\
        \Xcline{3-6}{0.4pt}
                              &       & DnCNN     &UNet          &ResUNet        &SwinIR          \\
        \Xhline{0.4pt}
          \multirow{2}*{MSE}  &  PSNR & 38.44     &39.01         & 39.51         &39.18       \\
                              &  SSIM & 0.9100    &0.9143        & 0.9187        &0.9143      \\
        \Xhline{0.4pt}
        \multirow{2}*{VIRNet} &  PSNR & 38.56     &39.28         & 39.63         &39.42   \\
                              &  SSIM & 0.9110    &0.9168        & 0.9194        &0.9166  \\
        \Xhline{0.4pt}
          \multicolumn{2}{c|}{\# Parameters (M)} 
                                      & 0.67      & 7.81         & 15.84         &11.50   \\
        \Xhline{0.4pt}
          \multicolumn{2}{c|}{\# FLOPs (G)} 
                                      & 175      & 168           & 658           &6017  \\
        \Xhline{0.8pt}
    \end{tabular}
\end{table}
\subsubsection{Hyper-parameter analysis}
Our method mainly involves two hyper-parameters, namely $\varepsilon_0$ in Eq.~\eqref{eq:prior-z} and $p$ in Eq.~\eqref{eq:xi}. We analyze the sensitiveness of our method to them on the task of real-world image denosing, since these two hyper-parameters both control the behavior of the noise modeling part in our method.  

\noindent \textbf{Hyper-parameter $\varepsilon_0^2$.}
As discussed in Sec.~\ref{subsucsec:discussion-elbo}, Our ELBO loss degenerates to MSE when $\varepsilon_0$ is set as an extremely small value. We thus trained the \textit{RNet} separately under MSE as a baseline for explicit comparison. The evaluative performance of VIRNet under different configurations of $\varepsilon_0$ is summarized in Table~\ref{tab:hyper_epsilon}. This table unveils the following observations: 1) Too large setting on $\varepsilon_0^2$ (e.g., 1e-4) yields a relatively inferior performance of VIRNet, attributed to the attenuation of the prior constraint on $\bm{z}$; 2) our VIRNet performs well as long as $\varepsilon_0^2$ resides within the range of [1e-5, 1e-7], and deteriorates obviously when $\varepsilon_0$ is smaller beyond the threshold of 1e-7; 3) VIRNet surpasses the MSE baseline by 0.12dB in terms of PSNR, indicating the pivotal role of noise modeling in our proposed methodology. Thus, we suggest to set $\varepsilon_0^2$ as 1e-6 in this work.

\noindent \textbf{Hyper-parameter $p$.}
In Eqs.~\eqref{eq:prior-sigma} and~\eqref{eq:xi}, a prior distribution is constructed for the noise variance $\sigma_i$, leveraging an initial value estimated within a $p \times p$ window. We present the quantitative comparison result of VIRNet under different settings for this hyper-parameter $p$ in Table~\ref{tab:hyper_p}. In general, the proposed VIRNet is insensitive to this hyper-parameter and showcases consistent performance across various configurations on $p$. That's because $\xi_i$ only provides a prior to constrain \textit{SNet}, and \textit{SNet} can be further adaptively rectified by the likelihood term in the ELBO loss during training. We simply set it as $7$ throughout all our experiments.

\subsubsection{Restoration Backbones} \label{subsubsec:network-ablation}
Our proposed method introduces a unified Bayesian framework to handle the task of blind IR, which does not rely on specific designs on the network architecture. By taking the limit as $\varepsilon_0 \rightarrow 0$, the framework then degenerates into the widely used MSE loss function. To comprehensively validate the superiority of our method beyond MSE, we consider four typically popular backbones, namely DnCNN~\cite{zhang2017beyond}, UNet~\cite{ronneberger2015u}, ResUNet~\cite{zhang2021plug}, and SwinIR~\cite{liang2021swinir}, for the restoration network \textit{RNet} in our method. We train these four backbones with MSE under the same settings to our method as baselines. Quantitative comparison on the task of real-world image denoising is itemized in Table~\ref{tab:backbone_ablation}. It can be seen that the proposed method brings up evident performance gain as compared with the MSE baseline, particularly for UNet and SwinIR, indicating the superiority of the proposed framework.

\section{Conclusion}
In this paper, we have proposed a novel deep variational network for blind IR, which aims to finely integrate the merits of both classical model-based methods and recent DL-based methods. On one hand, we have constructed a Bayesian generative model for blind image denoising and image super-resolution, by carefully considering the image degradation from the perspectives of image noise and blur kernel. On the other hand, a variational inference algorithm has been elaborately designed to solve the proposed model, in which the posteriori distribution is all parameterized by DNNs to increase the non-linear fitting capability. Most notably, this variational algorithm induces a unified framework to simultaneously deal with the tasks of degradation estimation and image restoration. Extensive experiments have also been conducted to demonstrate the superiority of our method on image denoising and super-resolution. In the future, we will make further efforts to extend our method to deal with more complicated and general image degradation. Besides, we will make efforts to explore more general kernel types beyond the Gaussian for SR and other image restoration tasks under our proposed statistical framework.

\section*{Acknowledgments}
This study is supported by the National Key R\&D Program of China (2020YFA0713900), the China NSFC projects under contracts 62076196, 62331028, 12226004.


\bibliography{reference}
%

%
\begin{IEEEbiography}[{\includegraphics[width=1in,height=1.25in,clip,keepaspectratio]{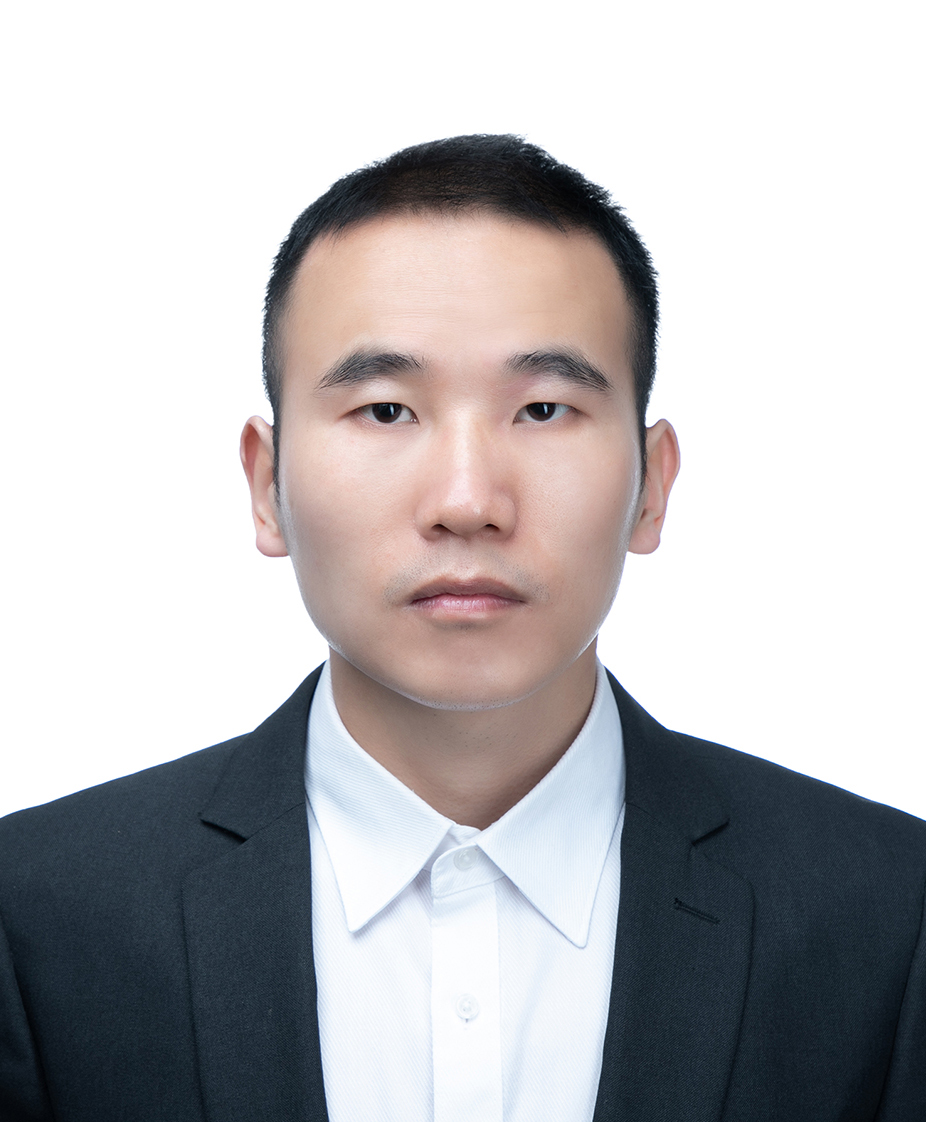}}]{Zongsheng Yue} (Member, IEEE) received his Ph.D. degree from Xi'an Jiaotong University, Xi'an, China, in 2021. He is currently a postdoctoral research fellow with school of Computer Science and Engineering, Nanyang Technological University. From September 2021 to March 2022, he was a associate research in the Department of Computer Science, Hong Kong University. He was a research assistant in the Department of Computing, Hong Kong Polytechnic University during October 2018 to June 2019 and the Institute of Future Cities, The Chinese University of Hong Kong during February 2017 to September 2017, respectively. His current research interests include noise modeling, image restoration, and diffusion model. 
\end{IEEEbiography}

\begin{IEEEbiography}[{\includegraphics[width=1in,height=1.25in,clip,keepaspectratio]{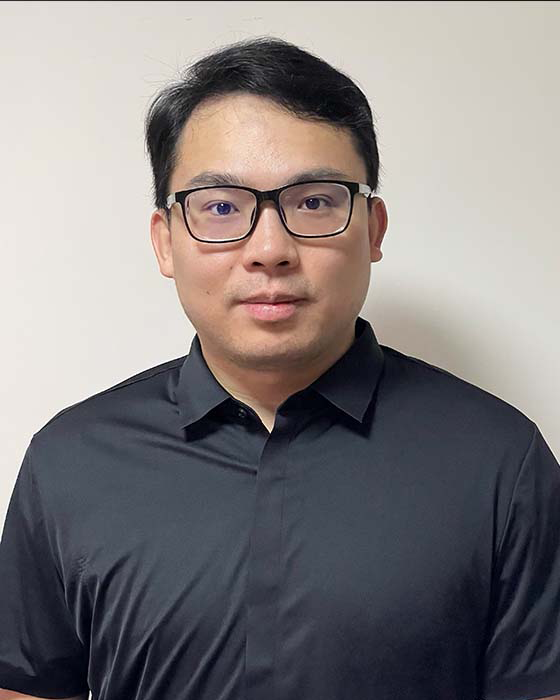}}]{Hongwei Yong} received his Ph.D. degree from the Department of Computing at the Hong Kong Polytechnic University in 2022 and received his MSc. degree and B.Sc degree from Xi’an Jiaotong University, China, in 2016 and 2013, respectively.  He is currently a research assistant professor in the Department of Computing, at the Hong Kong Polytechnic University (PolyU).  His research interests focus on the research areas of computer vision and deep learning optimization.  He has published a lot of research works in top journals and conferences including TPAMI, TIP, TNNLS, CVPR, ECCV, NeruIPS, etc. He is also a reviewer of some prestigious journals and conferences in my research areas, such as TPAMI, TIP, CVPR, AAAI, ICLR, NeruIPS, etc.
\end{IEEEbiography}

\begin{IEEEbiography}[{\includegraphics[width=1in,height=1.25in,clip,keepaspectratio]{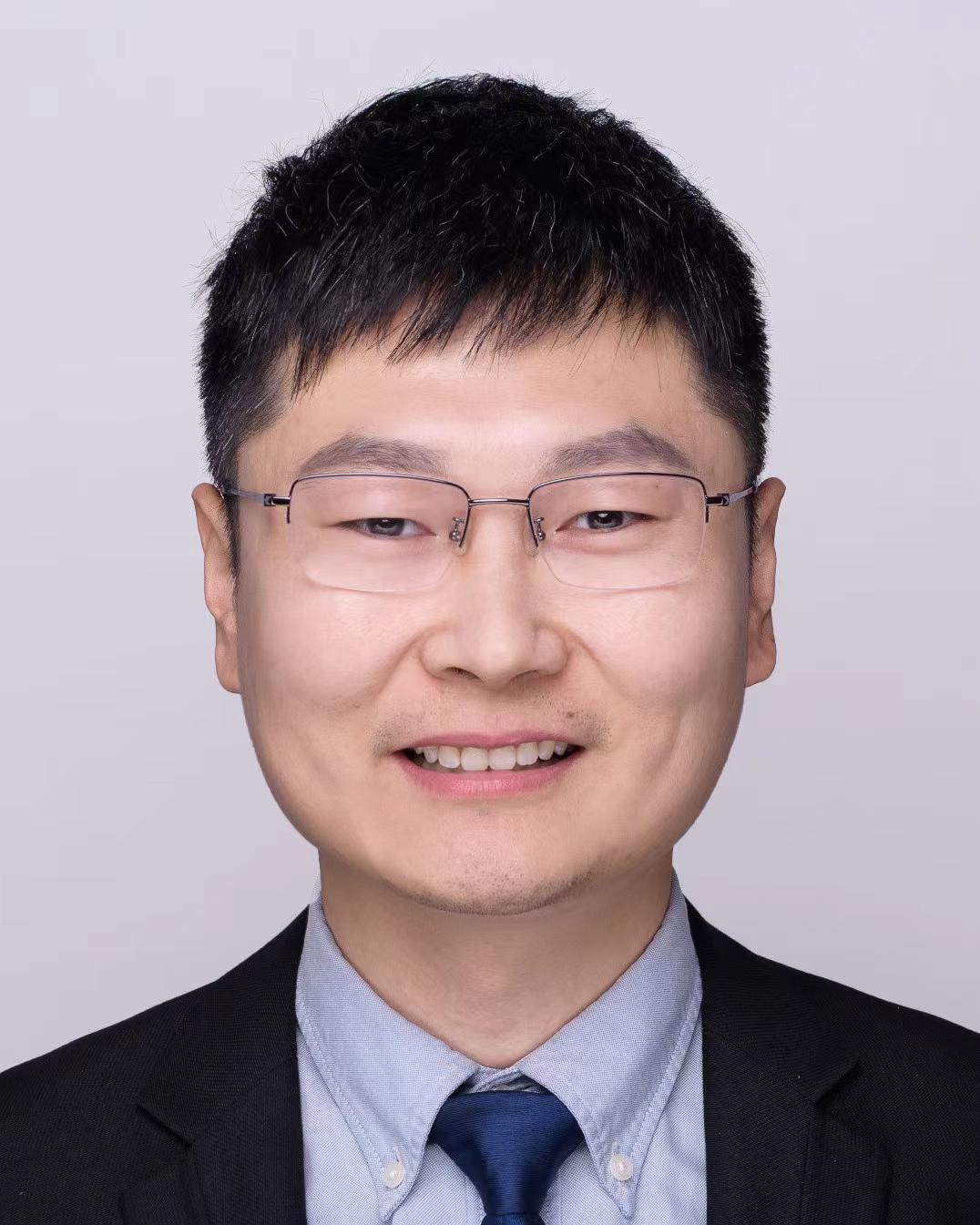}}]{Qian Zhao} (M'23) received the BSc and PhD degrees from Xi'an Jiaotong University, Xi'an, China, in 2009 and 2015, respectively. He was a visiting scholar with Carnegie Mellon University, Pittsburgh, PA, USA, from 2013 to 2014. He is currently an Associate Professor with the School of Mathematics and Statistics, Xi'an Jiaotong University. His current research interests include low-rank matrix/tensor analysis, Bayesian modeling and meta learning.
\end{IEEEbiography}

\begin{IEEEbiography}[{\includegraphics[width=1in,height=1.25in,clip,keepaspectratio]{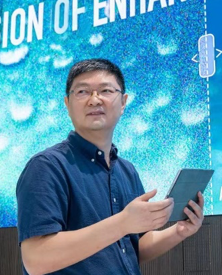}}]{Lei Zhang} joined the Department of Computing, The Hong Kong Polytechnic University, as an Assistant Professor in 2006. Since July 2017, he has been a Chair Professor in the same department. His research interests include Computer Vision, Pattern Recognition, Image and Video Analysis, Deep Learning, etc. Prof. Zhang has published more than 200 papers in those areas. As of 2023, his publications have been cited more than 90,000 times in literature. Prof. Zhang is a Senior Associate Editor of IEEE Trans. on Image Processing, and is/was an Associate Editor of IEEE Trans. on Pattern Analysis and Machine Intelligence, SIAM Journal of Imaging Sciences, IEEE Trans. on CSVT, and Image and Vision Computing, etc. He was listed as a “Clarivate Analytics Highly Cited Researcher” consecutively from 2015 to 2022. He is an IEEE Fellow. More information can be found in his homepage \url{http://www4.comp.polyu.edu.hk/~cslzhang/}.
\end{IEEEbiography}

\begin{IEEEbiography}[{\includegraphics[width=1in,height=1.25in,clip,keepaspectratio]{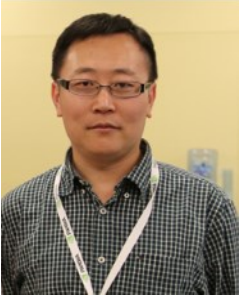}}]{Deyu Meng} received the B.Sc., M.Sc., and Ph.D. degrees from Xi'an Jiaotong University, Xi'an, China, in 2001, 2004, and 2008, respectively. He was a Visiting Scholar with Carnegie Mellon University, Pittsburgh, PA, USA, from 2012 to 2014. He is currently a professor with School of Mathematics and Statistics, Xi’an Jiaotong University, and adjunct professor with Faculty of Information Technology, The Macau University of Science and Technology. His current research interests include model-based deep learning, variational networks, and meta-learning.
\end{IEEEbiography}

\begin{IEEEbiography}[{\includegraphics[width=1in,height=1.25in,clip,keepaspectratio]{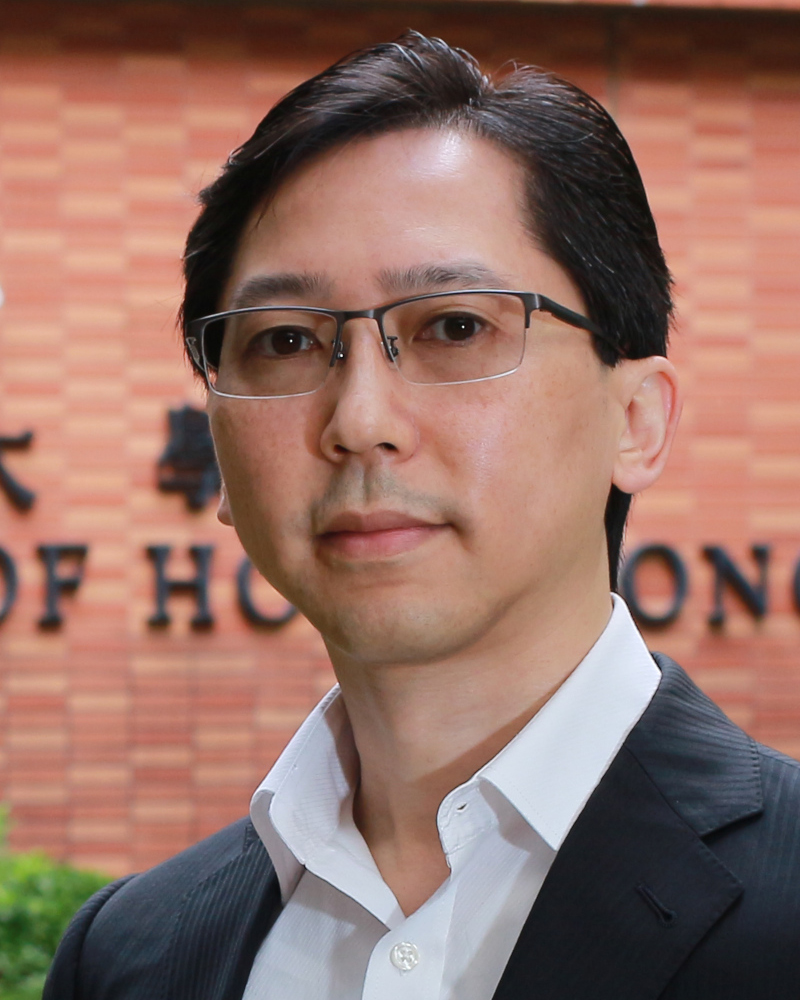}}] {Kwan-Yee K. Wong} (Senior Member, IEEE) received the B.Eng. degree (Hons.) in computer engineering from The Chinese University of Hong Kong in 1998, and the M.Phil. and Ph.D. degrees in computer vision (information engineering) from the University of Cambridge in 2000 and 2001, respectively. Since 2001, he has been with the Department of Computer Science at The University of Hong Kong, where he is currently an Associate Professor. His research interests include computer vision and machine intelligence. He is currently an Editorial Board Member of International Journal of Computer Vision (IJCV).
\end{IEEEbiography}

\clearpage
\onecolumn
\appendix

\subsection{Preliminaries on Inverse Gamma Distribution}\label{subsec:IGDistribution_supp}
The inverse Gamma distribution is a continuous statistical distribution on the positive real line,
whose probability density function is defined as follows:   
\begin{equation}
    f(x;\alpha,\beta) = \frac{\beta^{\alpha}}{\Gamma(\alpha)}\left(\frac{1}{x}\right)^{\alpha+1}\exp{\left(-\frac{\beta}{x}\right)},
    \label{eq:igamma_pdf}
\end{equation}
where $\Gamma(\cdot)$ denotes the Gamma function, $\alpha$ and $\beta$ are two parameters, called as ``shape'' and ``scale'', respectively. For the inverse Gamma distribution, its mode is $\frac{\beta}{\alpha+1}$.

\begin{figure}[h]
    \centering
    \includegraphics[scale=0.58]{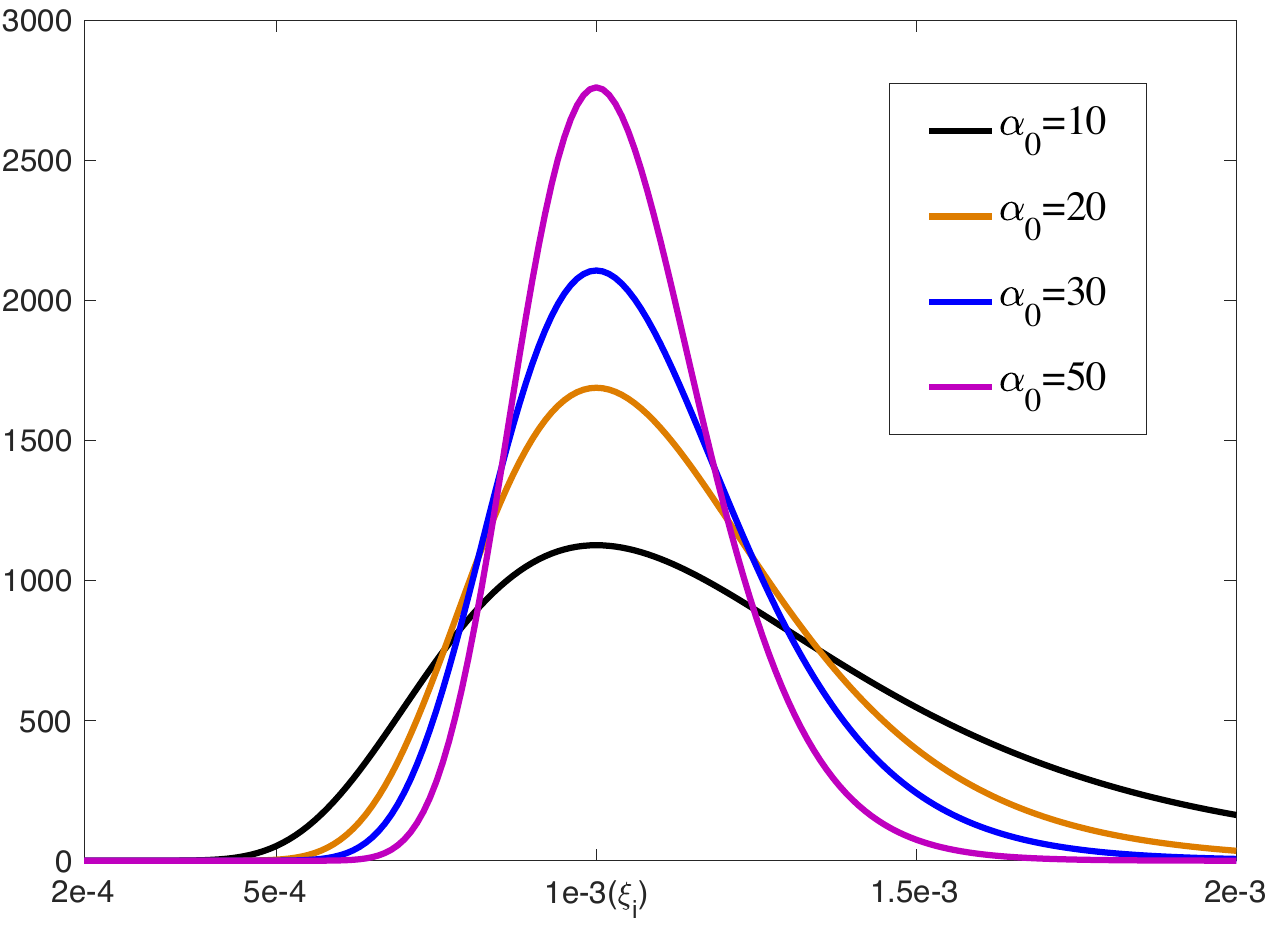}
    \caption{The probability density function of the inverse Gamma distribution $\text{IG}(\cdot|\alpha_0-1, \alpha_0*\xi_i)$
    under different values of $\alpha_0$ when $\xi_i=0.001$.}
    \label{fig:alpha_inv_supp}
\end{figure}

In Bayesian statistics, the inverse Gamma distribution is usually used as the conjugate prior for the unknown variance of a Gaussian distribution.
In this study, we also follow this principle and set the prior for the noise variance as inverse Gamma distribution in Eq.~\eqref{eq:prior-sigma} of the main text,
i.e.,
\begin{equation}
    \sigma_i^2 \sim \text{IG}\left(\sigma_i^2\bigg\vert\alpha_0-1, \alpha_0\xi_i\right), i=1,2,\cdots, d.
    \label{eq:prior-sigma-supp}
\end{equation}
With such settings for the shape and scale parameters, its mode just right equals $\xi_i$ that is an empirical estimation for the noise variance,
namely,
\begin{equation}
    \text{Mode}[\sigma_i^2] = \frac{\alpha_0\xi_i}{(\alpha_0-1)+1} = \xi_i.
    \label{eq:mode_sigma}
\end{equation}
Furthermore, the shape parameter $\alpha_0$ controls the strength of such prior constraint. Specifically, the larger $\alpha_0$ is,
the more accurate the estimated variance $\xi_i$ is. This can be intuitively explained by Fig.~\ref{fig:alpha_inv_supp}.

\vspace{2mm}
Similarly, for the blur kernel with covariance matrix
$\bm{\Sigma}=\begin{bmatrix}
    \lambda_1^2            & \lambda_1\lambda_2\rho \\ 
    \lambda_1\lambda_2\rho & \lambda_2^2   \\
\end{bmatrix}$, we also impose inverse Gamma distribution for $\lambda_1^2$ and $\lambda_2^2$ with shape parameter $\kappa_0$. Empirically, we find 
our method performs well and stably by setting $\kappa_0=50$ throughout all the experiments.

\newpage
\subsection{Details on Calculating the Evidence Lower Bound}\label{subsec:ELBO_supp}
In this section, we present the detailed calculations of some equations in the manuscript:
\begin{itemize}[itemsep=8pt]
    \item Calculation details of Eq.~\eqref{eq:marginal_likelihood}:
        \begin{align*}
            \log p(\bm{y})
            &= \int q(\bm{z},\bm{\sigma}^2,\bm{\varLambda}|\bm{y}) \log p(\bm{y}|\bm{z},\bm{\sigma}^2,\bm{\varLambda})
                \mathrm{d}\bm{z}\mathrm{d}\bm{\sigma} \mathrm{d}\bm{\varLambda}\\
            &= \int q(\bm{z},\bm{\sigma}^2,\bm{\varLambda}|\bm{y}) \log \left[
                \frac{p(\bm{y}|\bm{z},\bm{\sigma}^2,\bm{\varLambda})
                p(\bm{z})p(\bm{\sigma}^2)p(\bm{\varLambda})}{p(\bm{z},\bm{\sigma}^2,\bm{\varLambda}|\bm{y})}
                \right] \mathrm{d}\bm{z}\mathrm{d}\bm{\sigma}\mathrm{d}\bm{\varLambda} \\
            &= \int q(\bm{z},\bm{\sigma}^2,\bm{\varLambda}|\bm{y}) \log \left[
                \frac{p(\bm{y}|\bm{z},\bm{\sigma}^2,\bm{\varLambda})p(\bm{z})p(\bm{\sigma}^2)p(\bm{\varLambda})}
                {q(\bm{z},\bm{\sigma}^2,\bm{\varLambda}|\bm{y})}
                + \frac{q(\bm{z},\bm{\sigma}^2,\bm{\varLambda}|\bm{y})}{p(\bm{z},\bm{\sigma}^2,\bm{\varLambda}|\bm{y})}
                \right] \mathrm{d}\bm{z}\mathrm{d}\bm{\sigma}\mathrm{d}\bm{\varLambda}\\
            &= \int q(\bm{z},\bm{\sigma}^2,\bm{\varLambda}|\bm{y}) \log \left[
                \frac{p(\bm{y}|\bm{z},\bm{\sigma}^2,\bm{\varLambda})p(\bm{z})p(\bm{\sigma}^2)p(\bm{\varLambda})}
                {q(\bm{z},\bm{\sigma}^2,\bm{\varLambda}|\bm{y})}
                    \right] \mathrm{d}\bm{z}\mathrm{d}\bm{\sigma} \mathrm{d}\bm{\varLambda} \\
            &\mathrel{\phantom{=}} + \int q(\bm{z},\bm{\sigma}^2,\bm{\varLambda}|\bm{y}) \log \left[
                \frac{q(\bm{z},\bm{\sigma}^2,\bm{\varLambda}|\bm{y})}{p(\bm{z},\bm{\sigma}^2,\bm{\varLambda}|\bm{y})}
                \right] \mathrm{d}\bm{z}\mathrm{d}\bm{\sigma}\mathrm{d}\bm{\varLambda}\\
            &= E_{q(\bm{z},\bm{\sigma}^2,\bm{\varLambda}|\bm{y})}\Big[
                \log p(\bm{y}|\bm{z}, \bm{\sigma}^2,\bm{\varLambda})p(\bm{z})p(\bm{\sigma}^2)p(\bm{\varLambda})
                - \log q(\bm{z},\bm{\sigma}^2,\bm{\varLambda}|\bm{y})
                \Big] \\
            &\mathrel{\phantom{=}}+ KL\Big[q(\bm{z},\bm{\sigma}^2,\bm{\varLambda}|\bm{y}) \Vert p(\bm{z},\bm{\sigma}^2,\bm{\varLambda}|\bm{y})\Big].
        \end{align*}
    \item Calculation details of Eq.~\eqref{eq:ELBO_KL}:
        \begin{align*}
            \mathcal{L}(\bm{z},\bm{\sigma}^2,\bm{\varLambda};\bm{y})
                &= \int q(\bm{z},\bm{\sigma}^2,\bm{\varLambda}|\bm{y}) \log \left[
                    \frac{p(\bm{y}|\bm{z},\bm{\sigma}^2,\bm{\varLambda})p(\bm{z})p(\bm{\sigma}^2)p(\bm{\varLambda})}
                    {q(\bm{z},\bm{\sigma}^2,\bm{\varLambda}|\bm{y})}
                        \right] \mathrm{d}\bm{z}\mathrm{d}\bm{\sigma} \mathrm{d}\bm{\varLambda} \\
                &= \int q(\bm{z},\bm{\sigma}^2,\bm{\varLambda}|\bm{y}) \log \left[
                    \frac{p(\bm{y}|\bm{z},\bm{\sigma}^2,\bm{\varLambda})p(\bm{z})p(\bm{\sigma}^2)p(\bm{\varLambda})}
                    {q(\bm{z}|\bm{\sigma}^2,\bm{\varLambda},\bm{y})q(\bm{\sigma}^2|\bm{y})q(\bm{\varLambda}^2|\bm{y})}
                        \right] \mathrm{d}\bm{z}\mathrm{d}\bm{\sigma} \mathrm{d}\bm{\varLambda} \\
                &= \int q(\bm{z},\bm{\sigma}^2,\bm{\varLambda}|\bm{y})\Big[
                    \log p(\bm{y}|\bm{z},\bm{\sigma}^2,\bm{\varLambda})
                      \Big] \mathrm{d}\bm{z}\mathrm{d}\bm{\sigma} \mathrm{d}\bm{\varLambda}\\
                &\mathrel{\phantom{=}} + \int q(\bm{z},\bm{\sigma}^2,\bm{\varLambda}|\bm{y})\left[
                    \log\frac{p(\bm{z})}{q(\bm{z}|\bm{\sigma}^2,\bm{\varLambda},\bm{y})}
                      \right] \mathrm{d}\bm{z}\mathrm{d}\bm{\sigma} \mathrm{d}\bm{\varLambda}\\
                &\mathrel{\phantom{=}} + \int q(\bm{z},\bm{\sigma}^2,\bm{\varLambda}|\bm{y})
                    \log\frac{p(\bm{\sigma}^2)}{q(\bm{\sigma}^2|\bm{y})}
                       \mathrm{d}\bm{z}\mathrm{d}\bm{\sigma} \mathrm{d}\bm{\varLambda}\\
                &\mathrel{\phantom{=}} + \int q(\bm{z},\bm{\sigma}^2,\bm{\varLambda}|\bm{y})
                    \log\frac{p(\bm{\varLambda})}{q(\bm{\varLambda}|\bm{y})}
                       \mathrm{d}\bm{z}\mathrm{d}\bm{\sigma} \mathrm{d}\bm{\varLambda}\\
                &= E_{q(\bm{z},\bm{\sigma}^2,\bm{\varLambda}|\bm{y})}\Big[ \log p(\bm{y}|\bm{z},\bm{\sigma}^2,\bm{\varLambda}) \Big] \\
                &\mathrel{\phantom{=}} - E_{q(\bm{\sigma}^2,\bm{\varLambda}|\bm{y})} \bigg[
                    KL\Big[q(\bm{z}|\bm{\sigma}^2,\bm{\varLambda},\bm{y}) \Vert p(\bm{z})\Big]
                    \bigg] \\
                &\mathrel{\phantom{=}} - KL\Big[q(\bm{\sigma}^2|\bm{y}) \Vert p(\bm{\sigma}^2)\Big] \\
                &\mathrel{\phantom{=}} - KL\Big[q(\bm{\varLambda}|\bm{y}) \Vert p(\bm{\varLambda})\Big],
        \end{align*}
        where $q(\bm{\sigma}^2,\bm{\varLambda}|\bm{y})=q(\bm{\sigma}^2|\bm{y})q(\bm{\varLambda}|\bm{y})$.
    \item Calculation details of Eq.~\eqref{eq:likelihood_general}:
        \begin{align}
             E_{q(\bm{z},\bm{\sigma}^2,\bm{\varLambda}|\bm{y})}\left[ \log p(\bm{y}|\bm{z},\bm{\sigma}^2,\bm{\varLambda}) \right]
            & = \int q(\bm{z},\bm{\sigma}^2,\bm{\varLambda}|\bm{y})\log p(\bm{y}|\bm{z},\bm{\sigma}^2,\bm{\varLambda})
                \dif\bm{z}\dif\bm{\sigma}\dif\bm{\varLambda} \notag  \\
            & = \int q(\bm{z}|\bm{y},\bm{\sigma}^2,\bm{\varLambda})q(\bm{\sigma}^2|\bm{y})q(\bm{\varLambda}|\bm{y})
                \log p(\bm{y}|\bm{z},\bm{\sigma}^2,\bm{\varLambda})\dif\bm{z}\dif\bm{\sigma}\dif\bm{\varLambda} \notag\\
            &\approx \frac{1}{L} \sum_{j}^L \log p(\bm{y}|\tilde{\bm{z}}^{(j)},\tilde{\bm{\sigma}}^{(j)^2},\tilde{\bm{\varLambda}}^{(j)}) \notag \\
            & = \frac{1}{L} \sum_{j}^L \sum_i^d \log p(y_i|\tilde{\bm{z}}^{(j)},\tilde{\bm{\sigma}}^{(j)^2},\tilde{\bm{\varLambda}}^{(j)^2}) \notag  \\
            & = \frac{1}{L} \sum_{j}^L \sum_i^d \left\{
                -\frac{1}{2}\log 2\pi
                -\frac{1}{2}\log {\tilde{\sigma}^{(j)^2}}_i
                - \frac{\left(y_i-\left[\left(\tilde{\bm{z}}^{(j)} * \tilde{\bm{k}}^{(j)}\right)\Big\downarrow_s\right]_i\right)^2}{2\tilde{\sigma}^{(j)^2}_i}
                \right\},
        \end{align}
        where
        \begin{gather}
            \tilde{\bm{z}}^{(j)} = \bm{\mu} + \varepsilon_0 \bm{\epsilon}^{(j)},
            ~ ~ \bm{\epsilon}^{(j)} \sim \mathcal{N}(\bm{\epsilon}|0, \bm{I}_d), \label{eq:sample_z}\\ 
            \tilde{\bm{k}}^{(j)} = g\left(\tilde{\bm{\varLambda}}^{(j)}\right), ~ \tilde{\bm{\varLambda}}^{(j)} \sim q(\bm{\varLambda}|\bm{y}), \\
            \tilde{\bm{\sigma}}^{(j)^2} \sim q(\bm{\sigma}^2|\bm{y}). 
        \end{gather}
        After setting $L=1$, we can obtain the Eq.~\eqref{eq:likelihood_general} of the manuscript.
        \item The KL divergence of Gaussian and inverse Gamma distributions in Eqs.~\eqref{eq:KL-z}-\eqref{eq:KL_kernel} of the manuscript can be found in~\cite{wiki:normal} and \cite{wiki:inverse}, respectively.
\end{itemize}

\subsection{Additional Experimental Results on Image Supre-resolution} \label{subsec:exp-sr-supp}
In Table~\ref{tab:sisr_sf4} of the manuscript, we list the quantitative comparison results of different methods under scale factor 4 for the task of image super-resolution. Here we additionally present the results under scales 2 and 3 in Table~\ref{tab:sisr_sf2} and Table~\ref{tab:sisr_sf3}, respectively, for more comprehensive comparison. Obviously, compared with the results on scale factor 4 of the manuscript, the proposed method exhibits more evident superiorities under scale factors 2 and 3, which further demonstrates the effectiveness of our method.
\newpage
\begin{table*}[t]
    \centering
    \caption{Quantitative comparisons of different methods under scale factor 2. The PSNR/SSIM/LPIPS values in this table are all averaged
        over the seven kernels as shown in Fig. 5 of the manuscript. The results of the non-blind methods that rely on the pre-given ground truth blur
        kernel and noise level are marked in \textcolor[gray]{0.5}{gray} color to denote unfair comparisons.} \label{tab:sisr_sf2}
    \vspace{-4mm}
    \begin{tabular}{@{}C{3.6cm}@{}|@{}C{1.2cm}@{}|@{}C{1.40cm}@{}@{}C{1.45cm}@{}@{}C{1.45cm}@{}|
                                                  @{}C{1.40cm}@{}@{}C{1.45cm}@{}@{}C{1.45cm}@{}|
                                                  @{}C{1.40cm}@{}@{}C{1.45cm}@{}@{}C{1.45cm}@{}}
        \Xhline{0.8pt}
        \multirow{2}*{Methods} & \multirow{2}*{\makecell{Noise\\Level}} & \multicolumn{3}{c|}{Set14}
                                                                        & \multicolumn{3}{c|}{CBSD68}
                                                                        & \multicolumn{3}{c}{DIV2K100} \\
        \Xcline{3-11}{0.4pt}
                 &    & PSNR$\uparrow$  & SSIM$\uparrow$  & LPIPS$\downarrow$  & PSNR$\uparrow$  & SSIM$\uparrow$  & LPIPS$\downarrow$
                      & PSNR$\uparrow$  & SSIM$\uparrow$  & LPIPS$\downarrow$   \\
        \hline
        Bicubic  & \multirow{9}*{0.1}  & 28.04       & 0.7940       & 0.3096          & 27.76           & 0.7726      & 0.3745
                                     & 28.94       & 0.8195       & 0.3173               \\
        HAN~\cite{niu2020single}
                 &                   & 29.25       & 0.8301       & 0.2643          & 28.82           & 0.8108      & 0.3306
                                     & 30.23       & 0.8541       & 0.2738              \\
        IKC~\cite{gu2019blind}
                 &                   & 31.61       & 0.8868       & 0.1700          & 30.77           & 0.8711      & 0.2293
                                     & 32.57       & 0.9062       & 0.1782              \\
        DAN~\cite{huang2020unfolding}
                 &                   & 32.67       & 0.9048       & 0.1310          & 31.63           & 0.8939      & 0.1757
                                     & 33.36       & 0.9247       & 0.1319              \\
        DASR~\cite{wang2021unsupervised}
                 &                   & 32.19       & 0.8957       & 0.1586          & 31.40           & 0.8848      & 0.1994
                                     & 33.13       & 0.9167       & 0.1579              \\
        VIRNet (Ours)
                 &                   & 32.70       & 0.9047       & 0.1194          & 31.70           & 0.8936      & 0.1626
                                     & 33.61       & 0.9248       & 0.1193              \\
        \Xcline{3-11}{0.4pt}
        GT+SRMD~\cite{zhang2018learning}
                 &    & \textcolor[gray]{0.5}{32.85}           & \textcolor[gray]{0.5}{0.9041}         & \textcolor[gray]{0.5}{0.1348}
                      & \textcolor[gray]{0.5}{31.79}           & \textcolor[gray]{0.5}{0.8933}         & \textcolor[gray]{0.5}{0.1797}
                      & \textcolor[gray]{0.5}{33.87}           & \textcolor[gray]{0.5}{0.9233}         & \textcolor[gray]{0.5}{0.1404} \\
        GT+USRNet~\cite{zhang2020deep}
                 &    & \textcolor[gray]{0.5}{31.52}           & \textcolor[gray]{0.5}{0.9066}         & \textcolor[gray]{0.5}{0.1463}
                      & \textcolor[gray]{0.5}{30.45}           & \textcolor[gray]{0.5}{0.8964}         & \textcolor[gray]{0.5}{0.1930}
                      & \textcolor[gray]{0.5}{33.15}           & \textcolor[gray]{0.5}{0.9309}         & \textcolor[gray]{0.5}{0.1275}  \\
        \hline \hline
        Bicubic  & \multirow{9}*{2.55}  & 27.97           & 0.7872     & 0.3084     & 27.68     & 0.7653       & 0.3731
                                        & 28.82           & 0.8111     & 0.3192         \\
        DnCNN~\cite{zhang2017beyond}+HAN~\cite{niu2020single}
                 &                      & 28.95           & 0.8096     & 0.2565     & 28.54     & 0.7894       & 0.3352
                                        & 29.76           & 0.8288     & 0.2869         \\
        DnCNN~\cite{zhang2017beyond}+IKC~\cite{gu2019blind}
                 &                      & 29.35           & 0.8241     & 0.2459     & 28.99     & 0.8084       & 0.2992
                                        & 30.25           & 0.8479     & 0.2570         \\
        DnCNN~\cite{zhang2017beyond}+DAN~\cite{huang2020unfolding}
                 &                      & 30.01           & 0.8391     & 0.1946     & 29.55     & 0.8269       & 0.2402
                                        & 30.85           & 0.8626     & 0.2056         \\
        DASR~\cite{wang2021unsupervised}
                 &                      & 31.02           & 0.8620     & 0.2106     & 30.25     & 0.8442       & 0.2581
                                        & 31.80           & 0.8853     & 0.2133         \\
        VIRNet (Ours)
                 &                      & 31.48           & 0.8695     & 0.1764     & 30.64     & 0.8559       & 0.2198
                                        & 32.39           & 0.8952     & 0.1740         \\
        \Xcline{3-11}{0.4pt}
        GT+SRMD~\cite{zhang2018learning}
                 &       & \textcolor[gray]{0.5}{31.62}           & \textcolor[gray]{0.5}{0.8713}          & \textcolor[gray]{0.5}{0.1798}
                         & \textcolor[gray]{0.5}{30.62}           & \textcolor[gray]{0.5}{0.8554}          & \textcolor[gray]{0.5}{0.2274}
                         & \textcolor[gray]{0.5}{32.43}           & \textcolor[gray]{0.5}{0.8932}          & \textcolor[gray]{0.5}{0.1847} \\
        GT+USRNet~\cite{zhang2020deep}
                 &       & \textcolor[gray]{0.5}{31.08}           & \textcolor[gray]{0.5}{0.8798}          & \textcolor[gray]{0.5}{0.1833}
                         & \textcolor[gray]{0.5}{30.16}           & \textcolor[gray]{0.5}{0.8647}          & \textcolor[gray]{0.5}{0.2308}
                         & \textcolor[gray]{0.5}{32.40}           & \textcolor[gray]{0.5}{0.9033}          & \textcolor[gray]{0.5}{0.1741} \\
        \hline \hline
        Bicubic  & \multirow{9}*{7.65} & 27.41           & 0.7405          & 0.3892             & 27.13           & 0.7153         & 0.4845
                                       & 28.05           & 0.7548          & 0.4528                 \\
        DnCNN~\cite{zhang2017beyond}+HAN~\cite{niu2020single}
                 &                     & 27.15           & 0.6799          & 0.3880             & 26.77           & 0.6480         & 0.4854
                                       & 27.48           & 0.6719          & 0.4792                 \\
        DnCNN~\cite{zhang2017beyond}+IKC~\cite{gu2019blind}
                 &                     & 28.86           & 0.7985          & 0.2595             & 28.47           & 0.7803         & 0.3118
                                       & 29.60           & 0.8257          & 0.2725                 \\
        DnCNN~\cite{zhang2017beyond}+DAN~\cite{huang2020unfolding}
                 &                     & 29.01           & 0.8021          & 0.2374             & 28.57           & 0.7848         & 0.2854
                                       & 29.78           & 0.8299          & 0.2500                 \\
        DASR~\cite{wang2021unsupervised}
                 &                     & 29.81           & 0.8212          & 0.2528             & 29.02           & 0.7967         & 0.3133
                                       & 30.44           & 0.8475          & 0.2608                 \\
        VIRNet (Ours)
                 &                     & 30.03           & 0.8261          & 0.2277             & 29.24           & 0.8064         & 0.2807
                                       & 30.80           & 0.8558          & 0.2293                 \\
        \Xcline{3-11}{0.4pt}
        GT+SRMD~\cite{zhang2018learning}
                 &                     & \textcolor[gray]{0.5}{30.13}      & \textcolor[gray]{0.5}{0.8277}        & \textcolor[gray]{0.5}{0.2345}
                                       & \textcolor[gray]{0.5}{29.22}      & \textcolor[gray]{0.5}{0.8051}        & \textcolor[gray]{0.5}{0.2921}
                                       & \textcolor[gray]{0.5}{30.80}      & \textcolor[gray]{0.5}{0.8529}        & \textcolor[gray]{0.5}{0.2416} \\
        GT+USRNet~\cite{zhang2020deep}
                 &                     & \textcolor[gray]{0.5}{30.19}      & \textcolor[gray]{0.5}{0.8372}        & \textcolor[gray]{0.5}{0.2274}
                                       & \textcolor[gray]{0.5}{29.31}      & \textcolor[gray]{0.5}{0.8144}        & \textcolor[gray]{0.5}{0.2833}
                                       & \textcolor[gray]{0.5}{31.15}      & \textcolor[gray]{0.5}{0.8637}        & \textcolor[gray]{0.5}{0.2256} \\
        \Xhline{0.8pt}
    \end{tabular}
    \vspace{-2mm}
\end{table*}
\begin{table*}[t]
    \centering
    \caption{Quantitative comparisons of different methods under scale factor 3. The PSNR/SSIM/LPIPS values in this table are all averaged
        over the seven kernels as shown in Fig. 5 of the manuscript. The results of the non-blind methods that rely on the pre-given ground truth blur
        kernel and noise level are marked in \textcolor[gray]{0.5}{gray} color to denote unfair comparisons.} \label{tab:sisr_sf3}
    \vspace{-4mm}
    \begin{tabular}{@{}C{3.6cm}@{}|@{}C{1.2cm}@{}|@{}C{1.40cm}@{}@{}C{1.45cm}@{}@{}C{1.45cm}@{}|
                                                  @{}C{1.40cm}@{}@{}C{1.45cm}@{}@{}C{1.45cm}@{}|
                                                  @{}C{1.40cm}@{}@{}C{1.45cm}@{}@{}C{1.45cm}@{}}
        \Xhline{0.8pt}
        \multirow{2}*{Methods} & \multirow{2}*{\makecell{Noise\\Level}} & \multicolumn{3}{c|}{Set14}
                                                                        & \multicolumn{3}{c|}{CBSD68}
                                                                        & \multicolumn{3}{c}{DIV2K100} \\
        \Xcline{3-11}{0.4pt}
                 &    & PSNR$\uparrow$  & SSIM$\uparrow$  & LPIPS$\downarrow$  & PSNR$\uparrow$  & SSIM$\uparrow$  & LPIPS$\downarrow$
                      & PSNR$\uparrow$  & SSIM$\uparrow$  & LPIPS$\downarrow$   \\
        \hline
        Bicubic  & \multirow{9}*{0.1}  & 25.89       & 0.6988       & 0.4459          & 25.87           & 0.6765      & 0.5180
                                     & 26.78       & 0.7333       & 0.4513               \\
        HAN~\cite{niu2020single}
                 &                   & 26.82       & 0.7374       & 0.3884          & 26.63           & 0.7134      & 0.4606
                                     & 27.73       & 0.7702       & 0.3935              \\
        IKC~\cite{gu2019blind}
                 &                   & 27.76       & 0.7966       & 0.2805          & 27.46           & 0.7759      & 0.3408
                                     & 28.27       & 0.8249       & 0.2800              \\
        DAN~\cite{huang2020unfolding}
                 &                   & 29.08       & 0.8108       & 0.2699          & 28.42           & 0.7886      & 0.3297
                                     & 29.93       & 0.8436       & 0.2620              \\
        DASR~\cite{wang2021unsupervised}
                 &                   & 29.26       & 0.8147       & 0.2740          & 28.41           & 0.7911      & 0.3368
                                     & 29.49       & 0.8385       & 0.2783              \\
        VIRNet (Ours)
                 &                   & 29.44       & 0.8190       & 0.2518          & 28.66           & 0.7958      & 0.3148
                                     & 30.17       & 0.8467       & 0.2520              \\
        \Xcline{3-11}{0.4pt}
        GT+SRMD~\cite{zhang2018learning}
                 &    & \textcolor[gray]{0.5}{29.75}           & \textcolor[gray]{0.5}{0.8243}         & \textcolor[gray]{0.5}{0.2448}
                      & \textcolor[gray]{0.5}{28.82}           & \textcolor[gray]{0.5}{0.8005}         & \textcolor[gray]{0.5}{0.3099}
                      & \textcolor[gray]{0.5}{30.44}           & \textcolor[gray]{0.5}{0.8500}         & \textcolor[gray]{0.5}{0.2480} \\
        GT+USRNet~\cite{zhang2020deep}
                 &    & \textcolor[gray]{0.5}{29.28}           & \textcolor[gray]{0.5}{0.8332}         & \textcolor[gray]{0.5}{0.2426}
                      & \textcolor[gray]{0.5}{28.21}           & \textcolor[gray]{0.5}{0.8079}         & \textcolor[gray]{0.5}{0.3133}
                      & \textcolor[gray]{0.5}{30.37}           & \textcolor[gray]{0.5}{0.8647}         & \textcolor[gray]{0.5}{0.2295}  \\
        \hline \hline
        Bicubic  & \multirow{9}*{2.55}  & 25.84           & 0.6936     & 0.4637     & 25.81     & 0.6709       & 0.5330
                                        & 26.71           & 0.7268     & 0.4757         \\
        DnCNN~\cite{zhang2017beyond}+HAN~\cite{niu2020single}
                 &                      & 26.65           & 0.7207     & 0.4064     & 26.46     & 0.6959       & 0.4854
                                        & 27.47           & 0.7495     & 0.4361         \\
        DnCNN~\cite{zhang2017beyond}+IKC~\cite{gu2019blind}
                 &                      & 27.00           & 0.7362     & 0.3661     & 26.76     & 0.7139       & 0.4276
                                        & 27.80           & 0.7681     & 0.3728         \\
        DnCNN~\cite{zhang2017beyond}+DAN~\cite{huang2020unfolding}
                 &                      & 26.83           & 0.7292     & 0.3932     & 26.75     & 0.7105       & 0.4463
                                        & 27.75           & 0.7640     & 0.3943         \\
        DASR~\cite{wang2021unsupervised}
                 &                      & 28.54           & 0.7819     & 0.3029     & 27.79     & 0.7560       & 0.3689
                                        & 28.84           & 0.8099     & 0.3085         \\
        VIRNet (Ours)
                 &                      & 28.66           & 0.7843     & 0.2914     & 27.91     & 0.7568       & 0.3586
                                        & 29.20           & 0.8112     & 0.2961         \\
        \Xcline{3-11}{0.4pt}
        GT+SRMD~\cite{zhang2018learning}
                 &       & \textcolor[gray]{0.5}{28.90}           & \textcolor[gray]{0.5}{0.7893}          & \textcolor[gray]{0.5}{0.2859}
                         & \textcolor[gray]{0.5}{28.04}           & \textcolor[gray]{0.5}{0.7616}          & \textcolor[gray]{0.5}{0.3551}
                         & \textcolor[gray]{0.5}{29.49}           & \textcolor[gray]{0.5}{0.8168}          & \textcolor[gray]{0.5}{0.2905} \\
        GT+USRNet~\cite{zhang2020deep}
                 &       & \textcolor[gray]{0.5}{28.84}           & \textcolor[gray]{0.5}{0.8079}          & \textcolor[gray]{0.5}{0.2683}
                         & \textcolor[gray]{0.5}{27.84}           & \textcolor[gray]{0.5}{0.7789}          & \textcolor[gray]{0.5}{0.3397}
                         & \textcolor[gray]{0.5}{29.79}           & \textcolor[gray]{0.5}{0.8382}          & \textcolor[gray]{0.5}{0.2609} \\
        \hline \hline
        Bicubic  & \multirow{9}*{7.65} & 25.48           & 0.6577          & 0.5777             & 25.44           & 0.6328         & 0.6780
                                       & 26.21           & 0.6826          & 0.6491                 \\
        DnCNN~\cite{zhang2017beyond}+HAN~\cite{niu2020single}
                 &                     & 25.62           & 0.6239          & 0.5759             & 25.41           & 0.5934         & 0.6867
                                       & 26.12           & 0.6315          & 0.6582                 \\
        DnCNN~\cite{zhang2017beyond}+IKC~\cite{gu2019blind}
                 &                     & 26.51           & 0.7116          & 0.3652             & 26.20           & 0.6892         & 0.4283
                                       & 26.77           & 0.7419          & 0.3771                 \\
        DnCNN~\cite{zhang2017beyond}+DAN~\cite{huang2020unfolding}
                 &                     & 26.48           & 0.7066          & 0.4002             & 26.37           & 0.6846         & 0.4652
                                       & 27.33           & 0.7419          & 0.4097                 \\
        DASR~\cite{wang2021unsupervised}
                 &                     & 27.62           & 0.7415          & 0.3397             & 27.00           & 0.7123         & 0.4132
                                       & 28.10           & 0.7736          & 0.3472                 \\
        VIRNet (Ours)
                 &                     & 27.65           & 0.7431          & 0.3309             & 26.98           & 0.7121         & 0.4080
                                       & 28.12           & 0.7708          & 0.3430                 \\
        \Xcline{3-11}{0.4pt}
        GT+SRMD~\cite{zhang2018learning}
                 &                     & \textcolor[gray]{0.5}{27.75}      & \textcolor[gray]{0.5}{0.7449}        & \textcolor[gray]{0.5}{0.3337}
                                       & \textcolor[gray]{0.5}{27.05}      & \textcolor[gray]{0.5}{0.7144}        & \textcolor[gray]{0.5}{0.4143}
                                       & \textcolor[gray]{0.5}{28.31}      & \textcolor[gray]{0.5}{0.7751}        & \textcolor[gray]{0.5}{0.3447} \\
        GT+USRNet~\cite{zhang2020deep}
                 &                     & \textcolor[gray]{0.5}{28.24}      & \textcolor[gray]{0.5}{0.7690}        & \textcolor[gray]{0.5}{0.3010}
                                       & \textcolor[gray]{0.5}{27.35}      & \textcolor[gray]{0.5}{0.7361}        & \textcolor[gray]{0.5}{0.3796}
                                       & \textcolor[gray]{0.5}{28.95}      & \textcolor[gray]{0.5}{0.8000}        & \textcolor[gray]{0.5}{0.3040} \\
        \Xhline{0.8pt}
    \end{tabular}
    \vspace{-2mm}
\end{table*}

\end{document}